\documentclass[10pt]{article}

\usepackage{plain}

\usepackage{graphicx}
\usepackage{fullpage}

\usepackage{setspace}
\usepackage{epsfig}
\usepackage{amssymb}
\usepackage{amsmath}
\usepackage{subfigure}
\usepackage{wrapfig}
\usepackage{floatflt}
\usepackage[usenames]{color}
\usepackage[normalem]{ulem}
\usepackage{float}
\floatstyle{boxed}
\newfloat{boxtext}{t}{bxt}
\floatname{boxtext}{Box}

\newcommand{\comment}[1]{}

\begin{document}

\title{\bfseries\sffamily Bistability of an {\it In Vitro} Synthetic Autoregulatory Switch}

\author{ {\sf Pakpoom Subsoontorn$^1$, Jongmin Kim$^{1,4}$, Erik Winfree$^{2,3,4\hbox{\scriptsize *}}$}\\
Departments of {\bf \textsf{1}} Biology, {\bf \textsf{2}} Computation and Neural Systems, {\bf \textsf{3}} Computer Science \\and {\bf \textsf{4}} Bioengineering\\
California Institute of Technology\\
{\tt \{pakpoom,jongmin,winfree\}@dna.caltech.edu }}
\maketitle
\parbox{6.5in}{
\rule{6.5in}{.005in}   
\small\sffamily {\bfseries {\em Abstract:}}
\small\sffamily
The construction of synthetic biochemical circuits is an essential step for developing quantitative understanding of information processing in natural organisms.  Here, we report construction and analysis of an {\it in vitro} circuit with positive autoregulation that consists of just four synthetic DNA strands and three enzymes, bacteriophage T7 RNA polymerase, {\it Escherichia coli} ribonuclease (RNase)~H, and RNase~R.  The
modularity of the DNA switch template allowed a rational design of a synthetic DNA switch regulated by its RNA output acting as a transcription activator.  We verified that the thermodynamic and kinetic constraints dictated by the sequence design criteria were enough to experimentally achieve the intended dynamics: a transcription activator configured to regulate its own production.  Although only RNase~H is necessary to achieve bistability of switch states, RNase~R is necessary to maintain stable RNA signal levels and to control incomplete degradation products.   A simple mathematical model was used to fit ensemble parameters for the training set of experimental results and was then directly applied to predict time-courses of switch dynamics and sensitivity to parameter variations with reasonable agreement. The positive autoregulation switches can be used to provide constant input signals and store outputs of biochemical networks and are potentially useful for chemical control applications.
\rule{6.5in}{.005in}
 \\
}


\twocolumn

{\raggedleft \Large \bf Introduction} \\

Within a living cell lies an information processing system: the genetic circuits, the community of genes that regulate one another, and thus  allow the cell to express the right genes at the right times.
Synthetic biology provides a new approach to understand design principles underlying intricate and dynamic behaviors of
natural genetic circuits, by building and analyzing synthetic circuits which exhibit analogous behaviors;
this also lays the foundation for future engineering of complex chemical and biological systems.
With such circuits, it is possible to test hypotheses by construction, and often synthetic simplicity facilitates quantitative analysis as well as systematic engineering design~\cite{Endy2005, Andr2006, Benner2005, Hasty2001}.

For designing and constructing synthetic biochemical networks, several decision steps are necessary to choose the regulatory molecules and the biochemical infrastructure that supports network operation.
Protein-based synthetic circuits can take advantage of the huge diversity of protein structures and functions that allows a wide range of possible regulatory features~\cite{Benner2005,Andr2006,Elowitz00,Atkinson2003,Stricker2008}. Still, from an engineering perspective, it remains a challenge to rationally design a new regulatory protein with desirable function.  
RNA-based regulation is an alternative approach for controlling gene expressions ~\cite{Isaacs2004, Bayer2005, Isaacs2006, Win2007}.
RNA structures and interactions with other nucleic acid species can be reliably predicted based on Watson--Crick base-pairing, much more so than typical protein-protein or protein-DNA interactions of protein regulators.
As for the choice of biochemical infrastructure, the unintended interactions between the circuit and its environment can be greatly reduced by reconstructing the circuit \emph{in vitro}. \emph{In vitro} implementation of efficient transcription and translation machinery~\cite{Shimizu2001,jewett2008integrated} for synthetic networks have been successfully implemented~\cite{Noireaux2003}.
Yet, a supporting environment for an \emph{in vitro} RNA-based regulatory circuit can be even simpler as there is no need for translation, protein maturation, and protein-DNA interactions.

Previous work~\cite{Kim2004,Kim2006} introduced {\it in vitro} transcriptional circuits as simplified synthetic genetic regulatory circuits.  Individual switches functioning as inverters and a bistable feedback circuit composed of two inverters have been demonstrated (Figure~1A, top).
Our ``DNA switch", a simplified gene, has a promoter for T7 RNA polymerase (RNAP) flanked by two separate domains, an input domain and an output domain. Downstream of a promoter lies an output domain that encodes ``an RNA product"; on the opposite side of the promoter, an input domain regulated by ``an RNA regulator" via simple Watson-Crick base-pairing rule is located. The modularity of a DNA switch allows for an independent design of an RNA product and an RNA regulator within a switch. Hence, one can ``wire" several switches together to compose a complex regulatory network, in principle, by simply designing the RNA output of one switch to be the RNA regulator of the other switch.  Moreover, individual switch characteristics such as switching thresholds and maximum output levels are set by the concentrations of switch components rather than by molecular characteristics of binding domains. The state of each switch (transcription rate) and the levels of signals (RNA concentrations) relayed among switches define the behavior of the overall circuit.
As the circuit dynamics relies only on RNA transcription and degradation, our {\it in vitro} circuit operates in a relatively simple environment with NTP fuel and only a few enzymes, RNA polymerase and RNases.

Here, we expand the repertoire of circuit motifs for transcriptional circuits by implementing a repeater. A repeater is a transcriptional switch whose RNA output level is a sigmoidal activation function of its RNA regulator, thus providing a concise mechanism for relaying activation signal. An activator offers greater flexibility and simplicity for a synthetic circuit design and allows for faster timing.
Although, computationally, two inverters connected in a series can substitute for a repeater, such design can lead to delay and unintended amplification of noise~\cite{Hooshangi2005}.
In addition, a repeater wired to itself can implement a positive autoregulatory switch that simplifies our previous bistable circuit design~\cite{Kim2006} as demonstrated in synthetic {\it in vivo} circuits~\cite{Becskei2001,Isaacs2003} (Figure~1A, bottom).
In this report, we describe the design for a repeater switch and characterize each elementary reaction required for a functional repeater. We then construct and analyze a self-activating switch, a single repeater wired to itself, that can be tuned to exhibit bistability.  Furthermore, we characterize its sensitivity to parameters such as DNA and enzyme concentrations; we show that the experimental results are consistent with a simple mathematical model, whose parameters are set by Bayesian inference from a subset of experimental data.

\begin{figure*}[tbh!!!]
\centerline{\epsfig{figure=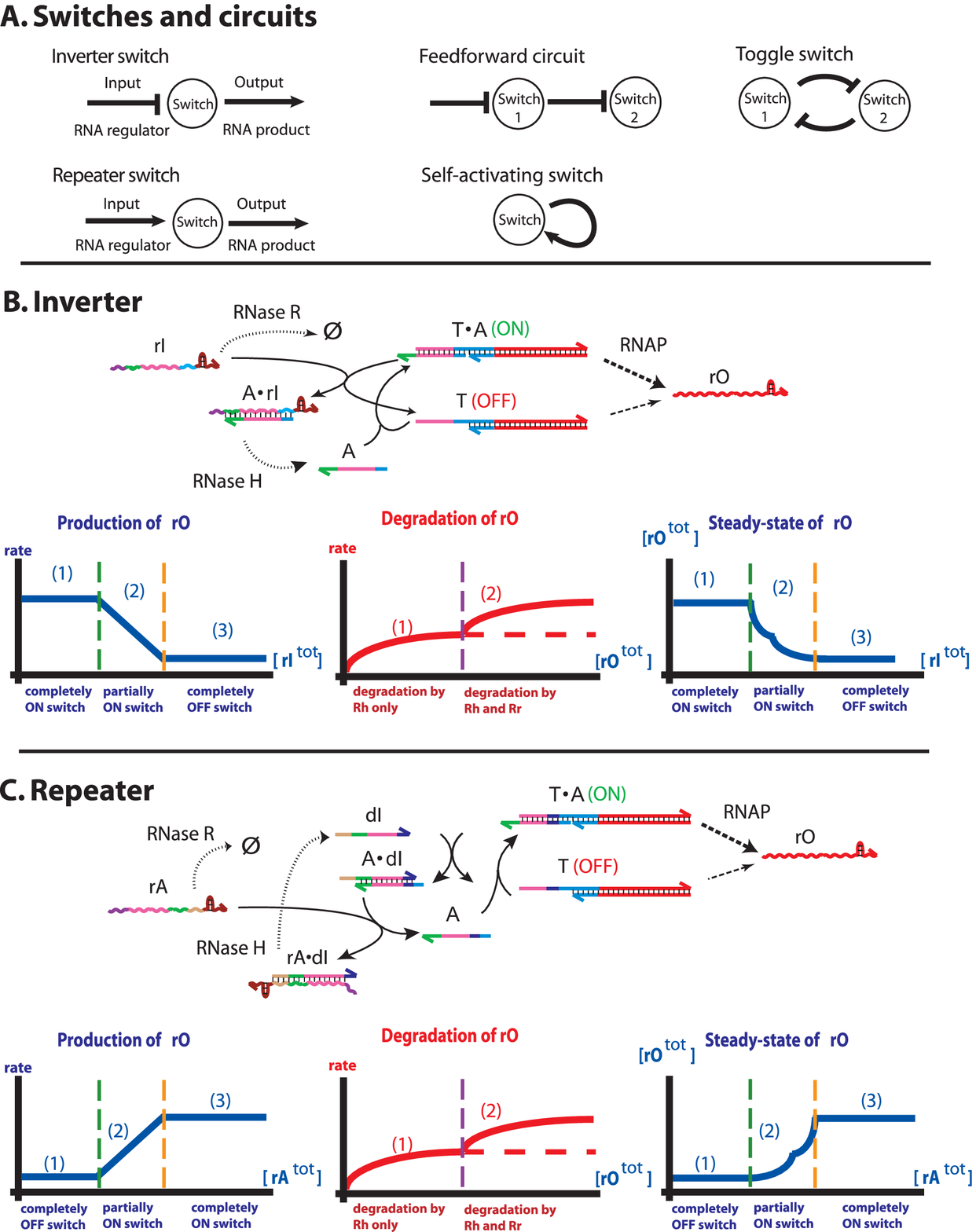, width = 0.85 \textwidth}}
\end{figure*}
\begin{figure*}
\caption{\textsf {Transcriptional switches and circuits.
(\textbf{A}) Schematic diagrams of an inverter, a repeater, and circuits. Blunt ends indicate inhibition, while arrowheads indicate signal production or activation.
(\textbf{B})  An inverter switch. T is the DNA template with an incomplete promoter region, A is the DNA activator, rI is the RNA inhibitor, and rO is the RNA output. Sequence domains are color-coded such that the same colors represent either complementary or identical sequences. Hybridization reactions are marked by black arrows. Transcription by RNAP and degradation by RNase~H and RNase~R are marked by black dashed arrows and dotted arrows, respectively. The production curve of rO as a function of rI, the degradation curve of rO as a function of rO, and the steady-state curve of rO as a function of rI constructed from the composition of the production and degradation curves are shown. Green dashed lines mark [rI${\rm ^{tot}}$] = [A${\rm ^{tot}}$] - [T${\rm ^{tot}}$], where all free A species are consumed by rI, yet the switch is still fully ON; orange dashed lines mark [rI${\rm ^{tot}}$] = [A${\rm ^{tot}}$], where all A species are consumed such that the switch is fully OFF. The purple dashed line marks [rO${\rm ^{tot}}$] = [dX${\rm ^{tot}}$] where dX is the regulatory target of rO: below this level rO is mostly bound to dX such that degradation curve is dominated by RNase~H, while above this level RNase~R degrades free rO. A dashed red line in degradation plot illustrates the case when only RNase~H is present in the reaction mixture. Note that the annihilation reaction between A and rI is not shown (see Figure~3).  Also, dX strand or RNase reactions on rO are not shown.
(\textbf{C})  A repeater switch. T and A are the same as described for an inverter switch, however, a repeater switch also uses a DNA inhibitor dI to set the switching threshold. Green dashed lines mark [rA${\rm ^{tot}}$] = [dI${\rm ^{tot}}$] - [A${\rm ^{tot}}$], where all free dI species are consumed by rA, yet the switch is fully OFF; orange dashed lines mark [rA${\rm ^{tot}}$] = [dI${\rm ^{tot}}$] - [A${\rm ^{tot}}$] + [T${\rm ^{tot}}$], where enough A species are freed from dI and available to hybridize with T, such that the switch is fully ON. The purple dashed line marks [rO${\rm ^{tot}}$] = [dX${\rm ^{tot}}$], where dX is the regulatory target of rO as above. Note that annihilation, interfering, recovering, recapturing reactions, dX strand, and RNase reactions on rO are not shown (see Figure~3).
 }}
\end{figure*}

\vspace{.1in}
\noindent
\textbf{Inverter versus Repeater}
\vspace{.1in}

The repeater that we implement in this paper shares several components and functional mechanisms with the previously implemented inverter.  Therefore, we first describe the components and functional mechanisms of a transcriptional inverter.  An inverter consists of two components, a DNA template (``T'') and a DNA activator (``A''). The DNA template T consists of single-stranded regulatory domain, partially single-stranded T7 RNAP promoter, and a double-stranded region  encoding an RNA output (``rO''). This partially single-stranded promoter is transcribed poorly by T7 RNA polymerase~\cite{Martin1987,Kim2006}, and thus, is designated as an OFF state.  The single-stranded DNA activator A is complementary to the missing promoter region of T. Upon hybridization of T and A, the resulting T$\cdot$A complex has a complete promoter except for a nick and was found to be transcribed well, approximately half as efficiently as a fully double-stranded promoter. Therefore, T$\cdot$A is designated as an ON state.
The inverter is regulated by an RNA inhibitor (``rI") that is complementary to A.
Because A$\cdot$rI complex is thermodynamically more stable than T$\cdot$A complex and the single-stranded domain of A beyond the helical domain of T$\cdot$A complex is available for initiating hybridization reaction with rI, rI can strip off A from T$\cdot$A complex through a toehold-mediated branch migration reaction~\cite{Mills1999,Seelig2005}.  Typically, A is in excess of T such that the input rI will first react with free A, then strip off A from ON-state switch, and the remaining rI will be free-floating in solution. Overall, the production rate, as well as the fraction of ON-state switch, is a sigmoidal inhibitory function of rI (Figure~1B, left).

The degradation speed of RNA signals plays an important role in circuit dynamics by setting the time-constants of signal relays. At the same time, the shape of the degradation curve in combination with the production curve would determine the steady-state RNA levels.
The degradation function is determined by the concentrations of substrates and the enzyme constants of RNases, {\it Escherichia coli} ribonuclease~H (RNase~H) and RNase~R.  RNase~H degrades RNA that is hybridized to DNA: rI in A$\cdot$rI complex and rO that is bound to its downstream regulatory target, ``dX'', to form a dX$\cdot$rO complex. At low rO concentrations, most of rO species exist within dX$\cdot$rO complex such that the degradation rate is largely dictated by RNase~H. On the other hand, RNase~R degrades single-stranded RNA, both free rI and free rO, such that the degradation rate of free rO --- the amount of rO in excess of total dX concentrations --- will be largely determined by RNase~R.
Because RNase~H has a low Michaelis constant, the degradation rate of rO by RNase~H quickly saturates as the concentration of dX$\cdot$rO increases. Also, the degradation rate of rO by RNase~R saturates as free rO concentration increases. Each degradation curve by RNase~H and RNase~R constitutes a typical Michaelis--Menten saturation curve with different origins: [rO${\rm^{tot}}$] = 0 for RNase~H and [rO${\rm^{tot}}$] = [dX${\rm^{tot}}$] for RNase~R. Thus, the composition of degradation curves by these two enzymes results in the degradation function with a kink located at the total concentration of dX  (Figure~1B, middle).  When the maximum rates and switching thresholds for production and degradation curves are approximately matched, the resulting steady-state RNA output level shows a sigmoidal inhibitory response with respect to RNA inputs, although the transition region contains a kink (Figure~1B, right).

A repeater utilizes some of the same modular design motifs as an inverter.
Because RNA polymerase transcribes poorly from a DNA/RNA hybrid promoter~\cite{McGinness2002}, we chose to implement a repeater through an indirect activation by an RNA activator. The transcriptional repeater also contains a DNA template T and a DNA activator A.  However, the repeater has an extra component, a DNA inhibitor (``dI''). Much like the RNA inhibitor rI of an inverter, the DNA inhibitor dI can bind to and remove A from the ON state T$\cdot$A complex and form a A$\cdot$dI complex.
The input of the repeater, an RNA activator, ``rA", is a single-stranded RNA that can displace dI within the inhibiting complex A$\cdot$dI and release A through a toehold-mediated branch migration. Then, the released A can bind back to T and turn the switch on.  Unlike the inverter, the concentrations of T and A are about the same and the concentration of dI is in excess of A to provide activation threshold for rA. (The DNA activator strand
concentration should be roughly comparable with the template
concentration; it should be at least as high, so that all the template
can be turned ON, but it needn't be higher, since excess activator
merely disables a stoichiometric amount of inhibitor.) The input rA will first react with free dI, then strip off dI from the A$\cdot$dI complex, and the remaining rA will be free-floating in solution. Overall, the production rate, as well as the fraction of ON-state switch, is a sigmoidal activation function of rA (Figure~1C, left).  The degradation function of a repeater is identical to that of an inverter (Figure~1C, middle). Therefore, the resulting steady-state RNA output level shows a sigmoidal activation response with respect to RNA input with a kink within the transition region (Figure~1C, right).

\begin{figure*}[tbh!!]
\centerline{\epsfig{figure=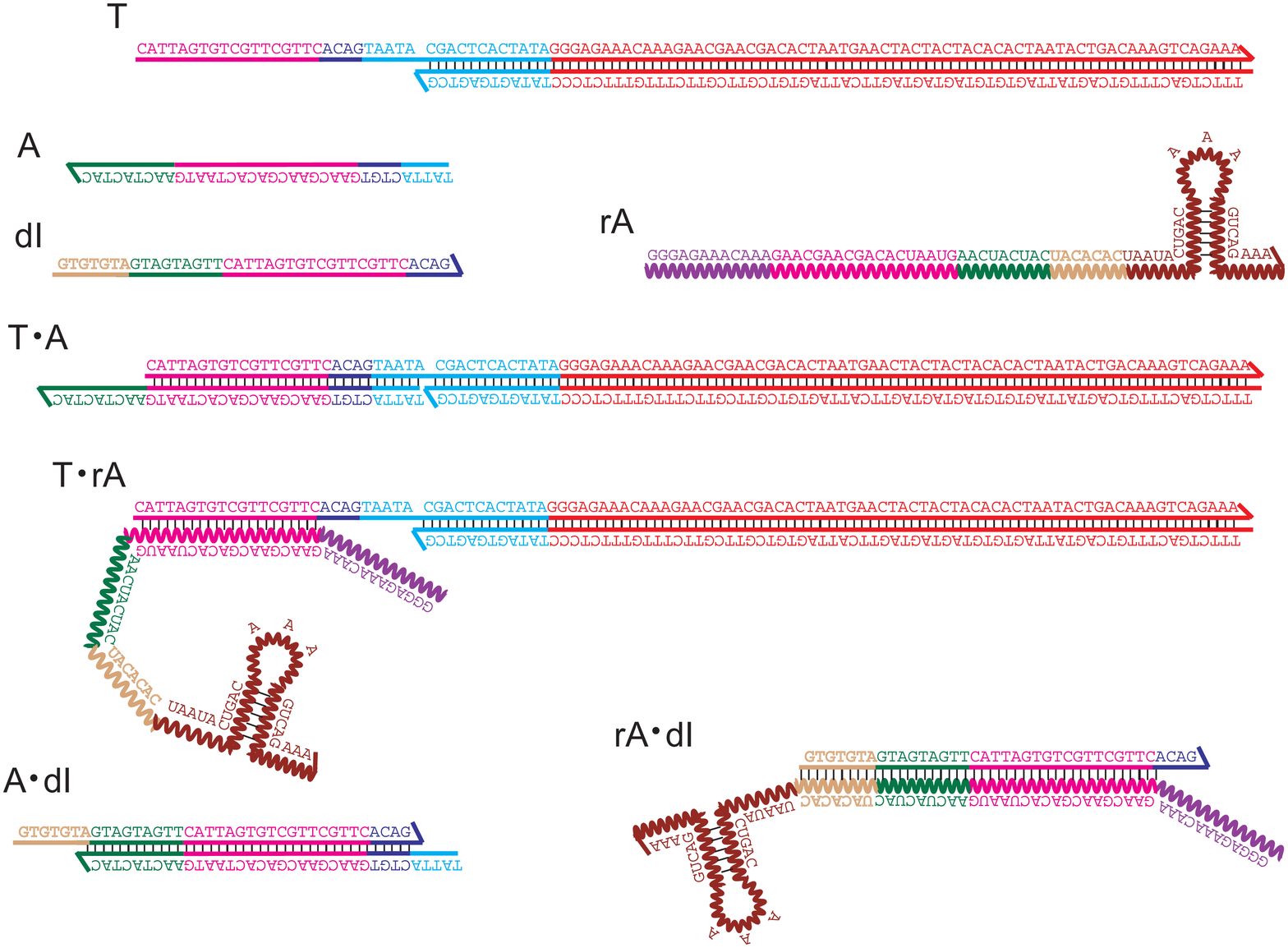, width = 0.9 \textwidth}}
\caption{\textsf{
Sequences of DNA and RNA species of a self-activating switch. The sequence domains are color-coded to indicate identical or complementary sequences.  Although we simply colored the output domain of switch coding for regulatory output ('rO' in Figure~1) red, this sequence is identical to the RNA input 'rA' because we constructed a self-activating switch.
 }}
\end{figure*}

\vspace{.3in}
\noindent{\Large \bf Results}\\
\vspace{.1in}
\noindent
\textbf{System Design and Mechanisms}
\vspace{.1in}

The challenge of designing sequences for a functional repeater lies in accommodating desirable hybridization reactions outlined above and suppressing side reactions and crosstalks. Here, we  present the sequence design for a repeater and provide evidence for its proper functionality.

Sequences of the repeater components are chosen to minimize
alternative folding~\cite{flamm00} and spurious
interactions~\cite{Seeman82}. Utilizing the modularity of synthetic switch designs, several domain lengths have been
adapted from previous work~\cite{Kim2006}; for example, the
binding domains of an OFF switch template to an activator (27 bases)
 and the toehold of an activator (9 bases).  The 3' end hairpin structure (16
bases) of RNA output increases copy number and also decreases self-coded extension of RNA transcripts by RNAP~\cite{Triana-alonso1995}.
The sequences of T, A, dI, rA, and their complexes are shown (Figure~2).
Note that, in this design, the sequence of RNA output, rO, is identical to RNA input, rA, because we constructed and characterized a self-activating switch.
Because we implemented an indirect activation by RNA inputs, the RNA activator rA and DNA activator A shares common sequence domains.  Thus, a T$\cdot$rA complex is expected to form when both T and rA are available.  However, it is not desirable if an excess of rA interferes with the hybridization reaction between T and A.
Thus, we implemented a staggered design, where the A$\cdot$dI complex leaves 5 bases of A as single-stranded (colored light blue in Figure~2) and the rA$\cdot$dI complex leaves 4 bases of dI as single-stranded (colored dark blue in Figure~2), resulting in a total of 9 base-pair differences between the proper ON state, T$\cdot$A, and the interfering complex, T$\cdot$rA. We chose the sequences and lengths of the binding domains for T, A, dI, and rA such that the predicted $\Delta G^{\circ}$ of complexes are in the following order: rA$\cdot$dI $<$ A$\cdot$dI $<$ T$\cdot$A $<$ T$\cdot$rA.
In terms of the number of base pairs, there are 18, 27, 31, and 34 complementary base-pairs for T$\cdot$rA, T$\cdot$A, A$\cdot$dI, and rA$\cdot$dI complexes, respectively.  Therefore, the RNA activator rA has the strongest influence on the state of a repeater by design. This design meets the requirement of proper hybridization reactions for a repeater as discussed below.

There are four types of simple hybridization reactions which we call activation, annihilation, subduing, and interfering (Figure 3A, red boxes):
A binds to T, forming T$\cdot$A (activation); dI binds to A, forming A$\cdot$dI (annihilation); rA binds to dI, forming rA$\cdot$dI (annihilation); rA binds to T, forming T$\cdot$rA (interfering).
Simple hybridization reactions are thermodynamically favorable because the resulting complex gains several base pairs, and hence the reactions proceed in a unidirectional way.  As mentioned above, `interfering' reaction is not desirable and we provide kinetic pathways (recovering and recapturing) to quickly resolve the interfering complex as shown below.
There are four types of strand displacement reactions, which we call inhibition, release, recovering, and recapturing (Figure 3A, orange boxes):
dI strips off A from T$\cdot$A complex to form A$\cdot$dI (inhibition); rA displaces A from A$\cdot$dI complex to form rA$\cdot$dI (release); A displaces rA from T$\cdot$rA complex to form T$\cdot$A (recovering); dI strips off rA from T$\cdot$rA complex to form rA$\cdot$dI (recapturing).
All displacement reactions are designed to be initiated at the `toehold,' a single-stranded overhang that extends beyond the helical domain of the initial complex. The incoming strand can bind to this toehold, providing a fast kinetic pathway through a toehold-mediated strand displacement reaction~\cite{Mills1999,Seelig2005}.
The predicted thermodynamic energies of starting complex and resulting complex indicate that all displacement reactions are thermodynamically favorable and hence approximately unidirectional.

\begin{figure*}[tbh!!!]
\centerline{\epsfig{figure=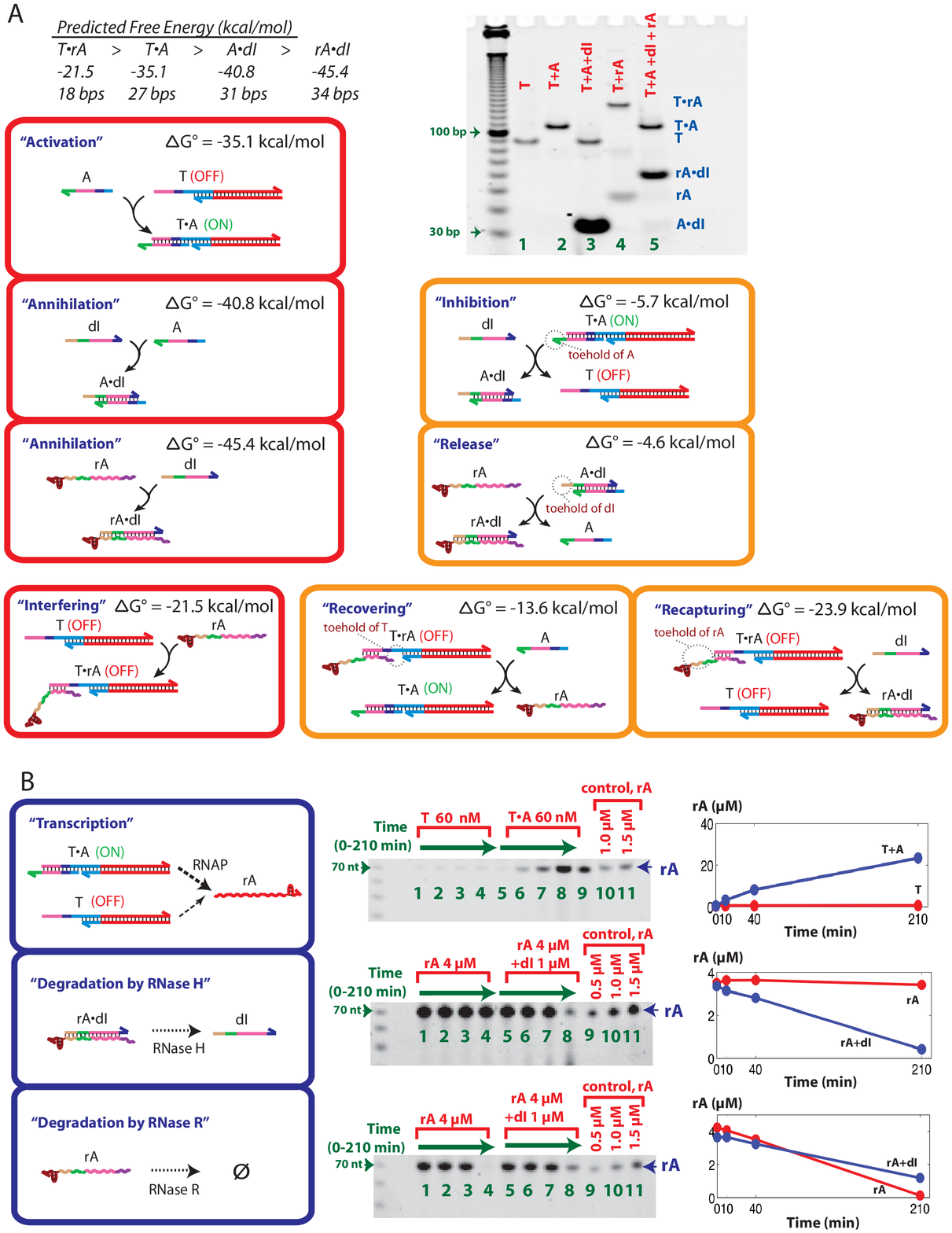, width = 0.85 \textwidth}}
\end{figure*}
\begin{figure*}
\caption{\textsf{Schematic representation of reactions for a repeater switch.
(\textbf{A})
DNA and RNA hybridization reactions. The red and orange boxed reactions show simple hybridization reactions and strand displacement reactions, respectively. Note that, all reactions are thermodynamically driven and that toehold-mediated branch migrations provide fast kinetic pathways for the strand displacement reactions. A non-denaturing gel was used to analyze the results of such hybridization and strand displacement reactions in the absence of enzymes (top right).
The DNA and RNA species were mixed and incubated for 5 min at room temperature prior to loading onto the non-denaturing gel. For each lane, the red label indicates the constituents loaded in that lane: T is 50 nM of [T${\rm ^{tot}}$]; A is 500 nM of [A${\rm ^{tot}}$]; dI is 700 nM of [dI${\rm ^{tot}}$]; and rA is 500 nM of [rA${\rm ^{tot}}$]. The leftmost lane contains a 10 base-pair ladder.
The blue labels on the right side of gel indicate corresponding single-stranded species and complexes. Single-stranded A and dI bands do not appear because they have been run off of the gel.
(\textbf{B}) Enzyme reactions. The reaction diagrams are shown in blue boxes with the corresponding denaturing gel analysis of experimental results on their right sides.
For the denaturing gels, the leftmost lanes contain 10 base ladders and the templates or substrates for enzymes are indicated by red labels on top of gels. Samples in lanes 1 through 4 (and also for lanes 5 through 8) were taken from  the same reaction tube at different time points (0, 10, 40, and 210 min). Sample in lane 9 of the gel analyzing RNAP reaction contained a half volume of that loaded in lane 8 to avoid SYBR gold signal saturation. Control lanes (lanes 10 and 11 for the gel analyzing RNAP reaction and lanes 9 through 11 for gels analyzing RNase~H and RNase~R reactions) have purified rA at the indicated concentrations. Enzyme concentrations used are as follows: [RNAP] = 33.4~nM, [RNaseH] = 0.168~nM, and [RNaseR] = 0.336~nM.}}
\end{figure*}

The hybridization reactions and strand displacement reactions for switch components are characterized by running different combinations of T, A, dI, and rA in a non-denaturing gel (Figure~3A).  The two single-stranded species comprising T was annealed prior to mixing with other single-stranded species.  The indicated components were simultaneously mixed in a test tube at room temperature and were allowed to sit for five minutes before being subjected to a non-denaturing gel run at 4$^\circ$C.
When T and A were mixed together, A could bind to T, resulting in a well-defined band of T$\cdot$A complex (lanes 1 vs.~2: activation).
When T, A, and an excess of dI were mixed together, only T and A$\cdot$dI complex were observed, implying that dI can bind to A and also strip off A from T$\cdot$A (lanes 2 vs.~3: annihilation and inhibition).
Of note, the single-stranded A and dI migrated faster than the 30-bp marker and were not visualized in the gel.
Also, when rA and T are mixed together, rA can bind to T and form a interfering complex T$\cdot$rA, which migrated slower than T$\cdot$A complex (lane 4: interfering).
However, when all four components were mixed together with the sum of A and rA concentrations in excess of dI, T$\cdot$rA complex was no longer visible, while T$\cdot$A and rA$\cdot$dI complexes were observed, implying that A and dI can strip off rA from T$\cdot$rA and dI can bind to rA (lanes 4 vs.~5: recovering, recapturing, and annihilation).
Moreover, it is also implied that rA can bind to dI and strip off dI from the A$\cdot$dI complex and release A, which in turn binds to T, resulting in a T$\cdot$A complex (lanes 3 vs.~5: annihilation, release, and activation).  We did not observe a noticeable A$\cdot$dI band in lane 5, indicating that the concentration of rA may have been underestimated for lane 5.
Recent reported measurements of DNA hybridization rates ranged from 10$^5$ to 10$^6$/M/s~\cite{Gao2006}.  Thus, starting from equal concentrations of switch components at hundreds of nM, the hybridization reactions would be completed with half-lives on the order of 10 to 100 seconds. No smearing or reaction intermediates were detected between the gel bands when several switch components were mixed together, indicating that the hybridization reactions and strand displacement reactions were mostly completed within a few minutes.

Three types of enzyme reactions are separately characterized as follows (Figure~3B).
RNAP could efficiently transcribe rA from an ON-state template, T$\cdot$A. However, the transcription was much slower from an OFF-state template, T (Figure~3B, top). Thus, transcription reaction is much more efficient from an ON-state switch than from an OFF-state switch as desired.  In this case, since dI was not provided in the transcription reaction, we do not expect to observe auto-regulation.
RNase~H could degrade rA when both rA and dI were present, but no degradation of rA was observed without dI (Figure~3B, middle). Thus, RNase~H degrades RNA that are hybridized with DNA, but not free-floating RNA.
RNase~R could degrade about 4~$\mu$M rA within 210 minutes. On the other hand, when 4~$\mu$M of rA was mixed with 1~$\mu$M of dI, about 1~$\mu$M of rA was left over after 210 minutes (figure~3B, bottom). Thus, RNase~R degrades single-stranded RNA, but not RNA within RNA-DNA hybrid complexes.

Taken together, our sequence design led to proper hybridization and strand displacement reactions among switch components with fast kinetics.  The enzyme reactions could provide production and degradation of RNA regulatory signals with specific recognition of substrates.
Also, note that the time-scales of enzyme reactions are much slower than typical hybridization or strand displacement reactions.
Therefore, the presumed sharp thresholds achieved by fast and irreversible hybridization kinetics would be approximately valid, even in the presence of constant enzyme-mediated production and degradation of signals.

\vspace{.1in}
\noindent
\textbf{Self-activating switch}
\vspace{.1in}

To demonstrate the functionality of the repeater mechanism, we chose to implement --- with a single self-regulating repeater switch ---  a bistable latch whose function is similar to that of the two-node network developed in~\cite{Kim2006}. The construction of a self-activating switch is straightforward as it only requires the RNA output rO to be identical to the RNA input rA, as shown in Figure 2.
Although we have shown that the elementary reaction steps are plausible, the repeater function embedded in a transcriptional circuit may show dynamic features deviating from our expectation.  For instance, previous works demonstrated that the switch behavior was measurably different when driven by an RNA species subject to constant production and degradation as compared to being driven by a DNA species~\cite{Kim2006}.  Therefore, the motivation is to assess how well the self-activating switch behaves with respect to its expected behavior and to gain quantitative understanding of activation switch motif within a system. Further, this simple circuit highlights the simplicity provided by the activation switch motif vis-a-vis creating the same functionality from a larger network of inhibitory switches.
In the following sections, we first `qualitatively' explain the dynamics of self-activating switch based on the production and degradation functions of RNA signals. Then, we assess how well the kinetic model using ensemble parameters matches the experimental results quantitatively.

\vspace{.1in}
\noindent
\textbf{Self-activating switch: qualitative and quantitative prediction of bistability}
\vspace{.1in}

For a self-activating switch, the RNA input rA and RNA output rO are identical, and therefore, we call RNA output as `rA' henceforth. The target DNA of the RNA output, dX, is now the DNA inhibitor, dI. Therefore, the production and degradation curves shown in Figure 1C can now be interpreted as the production and degradation curves of RNA activator rA as functions of its own concentrations. Thus, given the same switch template and DNA activator concentrations ([T${\rm^{tot}}$] = [A${\rm^{tot}}$]), the concentration of rA that consumes all dI and turn the switch ON in the production curve coincides with the threshold in the degradation curve below which degradation by RNase~H is dominant (i.e.~[rA${\rm^{tot}}$] = [dI${\rm^{tot}}$]). The intersections of production and degradation curves thus indicate the steady-states where total rA concentrations remain constant, and where and how they cross will dictate whether the self-activating switch show monostable or bistable behavior. If an unstable fixed point exists where a slight increase in rA drives further rA production and a slight decrease in rA drives further rA reduction (Figure~4A, right), the system will show bistable behavior with the threshold set at this unstable fixed point.  In that case, the switch states approach either of the two stable steady-states, completely ON or OFF, depending on the initial concentration of rA.
On the other hand, a monostable ON state will be achieved irrespective of initial RNA activator concentration if the production rate exceeds the degradation rate at the switching threshold (Figure~4A, left); a monostable OFF state will be achieved irrespective of initial RNA activator concentration if the degradation rate exceeds the production rate of the RNA activator when the switch is fully ON (Figure~4A, center).  The bistable and monostable switch behaviors depend not so much on the physical nature of molecular components but on the features set by continuously tunable concentrations --- such as the output amplitude and the activation threshold.

\begin{figure*}[tbh]
\centerline{\epsfig{figure=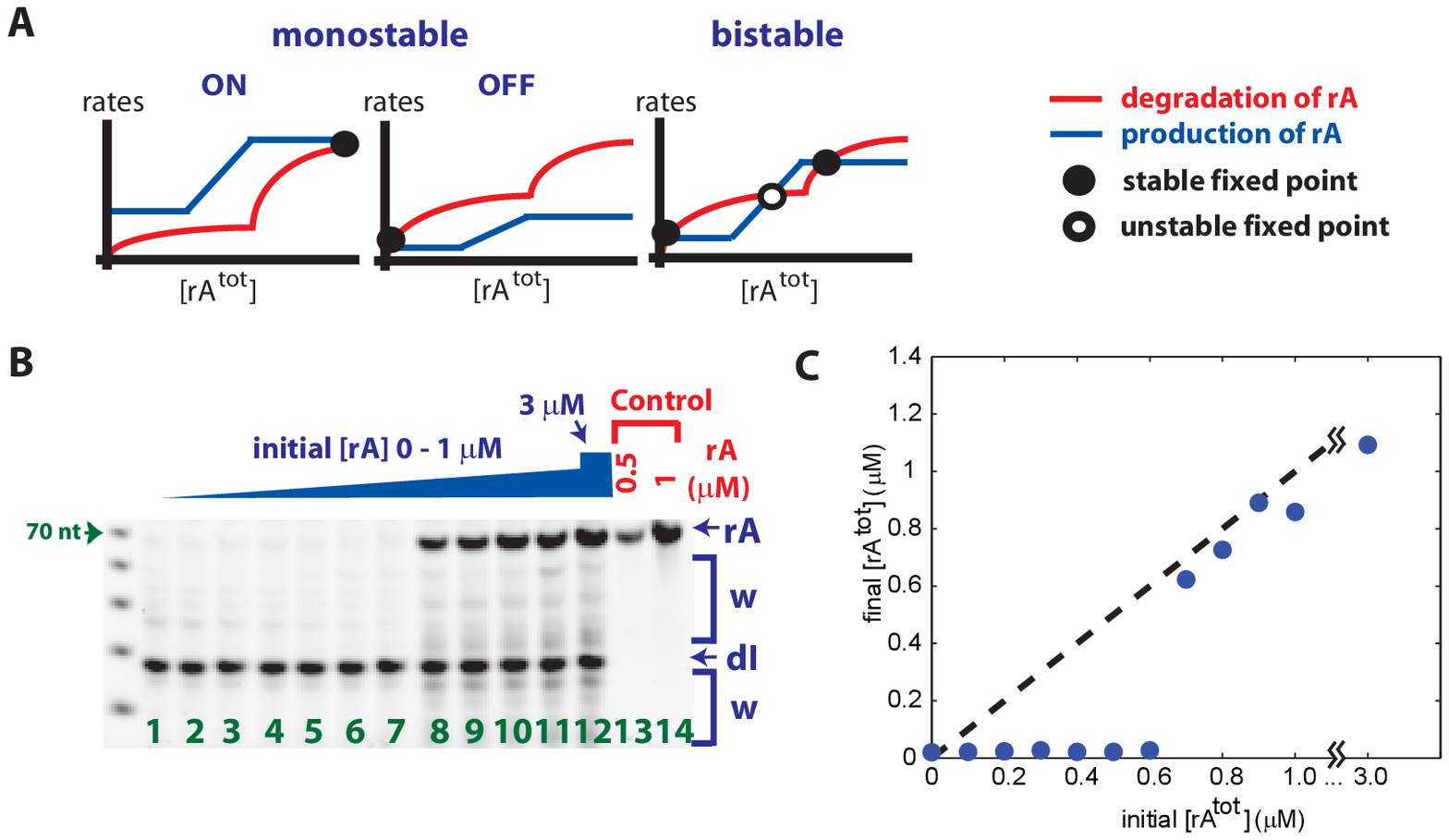, width = 0.95 \textwidth}}
\caption{\textsf{Bistability of a self-activating switch. (\textbf{A}) Production and degradation functions, and dynamics of a self-activating switch. Three plots illustrate possible configurations of production and degradation functions of rA as a function of rA. Depending on the locations and stabilities of fixed points, two types of monostable behaviors or a bistable switch behavior are expected.
(\textbf{B}) A denaturing gel analysis to test bistability of a self-activating switch. Lanes 1 through 12 show the results from 12 separate reactions. The reaction conditions are as follows: [T${\rm ^{tot}}$]  = 50 nM, [A${\rm ^{tot}}$]  = 90 nM, [dI${\rm ^{tot}}$]  = 1000 nM, [RNAP] = 66 nM, [RNaseH]  = 0.7 nM, and [RNaseR] = 0.23 nM, with variable initial rA concentrations ([rA${\rm ^{tot}}$] = 0, 0.1, 0.2, ..., 1, and 3 $\mu$M).  After 120 min, all reactions were stopped and subject to denaturing gel analysis. Control lanes, lanes 13 and 14, have purified rA at the indicated concentrations.  Brackets marked `w' indicate incomplete transcription products and degradation products.
(\textbf{C}) The measured final concentrations of rA from the denaturing gel in (B). Note that the apparent threshold of switching between the low and high steady-states is between 0.6 and 0.7 $\mu$M of initial rA concentrations. Black dashed line marks the points where the initial and final rA concentrations coincide.
}}
\end{figure*}

We then experimentally tested whether the self-acti\-vating switch can show bistable output responses (Figure~4B). When the self-activating switch was subject to transcription and degradation reactions with a wide range of initial RNA activator concentrations, the final RNA activator concentrations reached two distinct states after two hours: almost 0 nM or higher than 600 nM. The self-activating switch showed bistable behavior as expected.
However, the experimental results deviated from an idealized circuit behavior in two ways: a low switching threshold and non-constant ON state outputs. First, the switching threshold lied between 600 and 700 nM of rA in the experiment while the expected threshold from idealized production and degradation curves lied between 910 and 960 nM of rA for the case shown in Figure~4B. This discrepancy may be explained by the ``burst phase'' in enzyme
kinetics~\cite{Jia1997}, which could add a few copy-numbers of RNA transcripts rA and effectively lowered the apparent switching threshold for initial rA concentrations.
Second, the ON state output levels were not constant in contrast to almost identical OFF state output levels. High final RNA outputs were measured for high initial RNA concentrations, which could be explained by a slow degradation kinetics near the steady-state. On the other hand, the measured RNA output levels were as low as 600 nM when more than 960 nM of rA would be required to maintain a fully ON state in the idealized model.
We used RNase~R to clean up the incomplete single-stranded degradation products by RNase~H. Nevertheless, incomplete degradation products accumulated over time (Figure~4B, brackets marked `w'). Thus, it is possible that some portions of these incomplete degradation products still function much like a full-length RNA activator --- e.g. rA without the hairpin region at the 3' end.

Due to experimental difficulties including RNase~R not being commercially available at the time of experiment and having apparently short life-time, we focused our study on a simpler system without RNase~R. Further, excluding RNase~R eliminates the kink within degradation function and possibly allow for a simpler quantitative interpretation.  However, since the degradation curve by RNase~H is insensitive to the change of rA concentrations when the concentration of rA is above the total concentration of dI, the production and degradation curves may not cross at the high steady-state. In that scenario, the high state of rA concentration can grow without bound --- which make the system behavior more latch-like.  Even so, the behavior of switch state is bistable.

 \begin{figure*}[tbh!!]
\centerline{\epsfig{figure=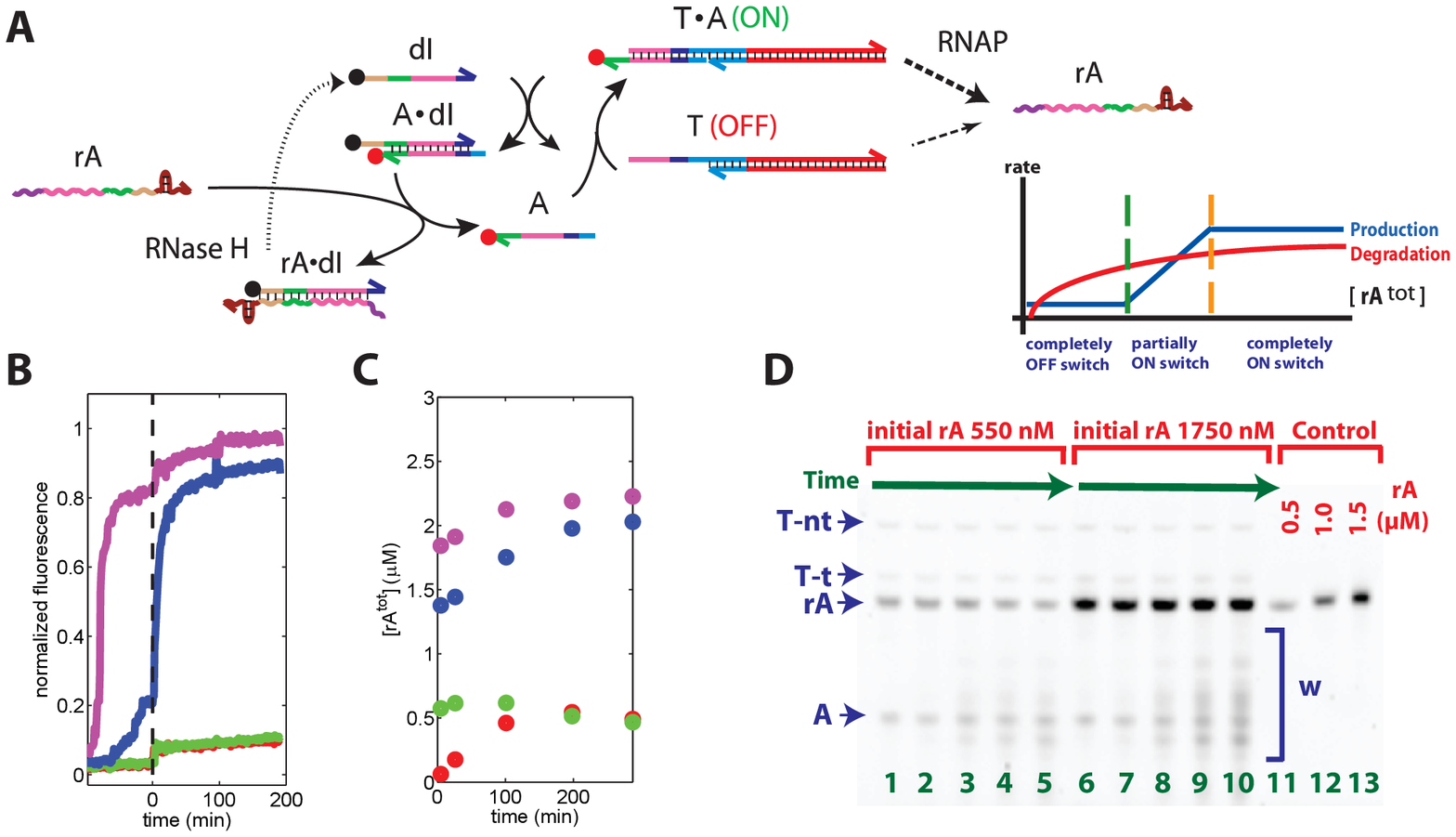, width = 0.95 \textwidth}}
\caption{\textsf{Time courses of a self-activating switch.
(\textbf{A})
The schematic reaction diagram of a self-activating switch (cf.~Figure~1C). The activator A is labeled with Cy5 fluorophore (red circle) and the inhibitor dI is labeled with IowaBlack-RQ quencher (black circle) for real-time monitoring of switch states by fluorescence. A possible arrangement of production and degradation curves for self-activating switch without RNase~R in the reaction mixture is shown as an inset (cf.~Figure~4A). (\textbf{B}) For fluorescence measurements, all the switch components, DNA template T, fluorophore-labeled A, and quencher-labeled dI, were added to each cuvette. Then, purified rA was added prior to enzyme additions such that the initial conditions are different for each cuvette. Finally, the enzymes were added at time 0 and thoroughly mixed (black dashed line). Normalized fluorescence signal measures the fraction of A that is not bound to dI, i.e.~([A]+[T$\cdot$A])/[A${\rm ^{tot}}$], because the fluorescence from A is quenched when A is in a A$\cdot$dI complex. All four cuvettes contained 48~nM of [T${\rm ^{tot}}$], 145~nM of [A${\rm ^{tot}}$], 1500~nM of [dI${\rm ^{tot}}$], 16.7~nM of [RNAP], and 1.68~nM of [RNaseH] with the initial rA concentrations at 0, 550, 1350, and 1750~nM (colored red, green, blue, and magenta, respectively).
(\textbf{C}) Time courses of [rA${\rm^{tot}}$] with samples taken from four different cuvettes. The colors correspond to those of fluorescence traces in (B).
(\textbf{D}) One of the denaturing gels for [rA${\rm^{tot}}$] measurements as shown in (C). Samples in lanes 1 through 5 (as well as lanes 6 through 10) were taken from the same cuvette at different time points. The final rA concentrations were measured with respect to the known concentrations of purified rA in the control lanes, lanes 11 through 13.
}}
\end{figure*}

To test bistability of switch response in the absence of RNase~R, we monitored the kinetics of switch response through real-time fluorescence measurement.  The 3' end of A is labeled with a Cy5 dye and the 5' end of dI is labeled with an IowaBlack-RQ quencher such that the fluorescence is low when the A$\cdot$dI complex is formed due to fluorescence quenching~\cite{Marras2002} while the fluorescence is high when A is free or within the T$\cdot$A complex (Figure~5A). Of note, fluorophore-quencher interaction can stabilize the resulting complex, A$\cdot$dI~\cite{Marras2002}. (In preliminary work, we observed mild to severe differences between fluoro\-phore-labeled strands and their unlabeled twins, depending on the fluorophore used.)  We initiated the reaction with four different rA concentrations and measured how the system approached different steady-states (Figure~5B). The fluorescence traces either reached a maximum signal, which implied a completely ON state, or a minimum signal, which implied a completely OFF state. The fluorescence monitoring indicated that the self-activating switch quickly approached steady-states and that both fully ON and OFF cases were stable steady-states. At the same time, the dynamics of RNA signal was determined by taking samples at different time points and measuring the total rA concentrations in gel (Figure~5CD). The high fluorescence traces (magenta and blue) corresponded to more than 2~$\mu$M final RNA levels, but the low fluorescence traces (green and red) corresponded to about 0.5~$\mu$M final RNA levels. Thus, we achieved bistable behavior in both switch and RNA signals, albeit the fact that the high RNA output level is not bounded.

While the bistable system behavior was demonstrated experimentally, a mathematical framework for the {\it in vitro} self-activation system would be required to investigate several further questions.
First, do the elementary reactions prescribed in Figure~3 explain the system behavior? Second, how fine-tuned the rate constants have to be to achieve bistability? Third, are the estimates and bounds for rates obtained from the characterization of elementary reactions compatible with the system behavior?
To address these questions, we constructed a simple mathematical model for the self-activating switch that uses four ordinary differential equations. The dynamics of each DNA and RNA species were derived from the hybridization and strand displacement reactions described in Figure~3, assuming Michaelis--Menten enzyme kinetics (see Appendix). Note that the concentration of `interfering complex' T$\cdot$rA was assumed to be negligible in the simple model, and thus, `interfering', `recovering', and `recapturing' reactions were not included. Initial simulation results showed that a bistable switch response can be achieved with plausible rate parameters similar to a previous work~\cite{Kim2006} (data not shown).

To test whether our mathematical model can be easily generalized using a relatively small number of experiments to determine system behavior for a wider range of parameter choices, we need a separate training and test data sets.
Yet, a rich enough training data set would be required to constrain parameters effectively. Thus, the training data set was obtained by measuring final rA concentrations for self-activating switches initialized with a wide range of rA concentrations and three different dI concentrations as switching thresholds (Figure~6A, red, blue, and green).  For this data set, fluorescence data for available A concentrations were not measured.  Below the apparent switching threshold, the final concentrations of rA approached $\sim$0.5~$\mu$M regardless of the initial rA concentrations. On other other hand, for initial rA concentrations higher than the apparent switching thresholds, the final rA concentrations were distinctively higher. As expected from the analysis of idealized production and degradation curves shown in Figure~5A, adjusting dI concentrations shifted the threshold of switching accordingly.

To find suitable values for the set of 11 parameters in our model, we used a
Monte Carlo Bayesian inference approach that results in an ensemble of
parameter sets that are compatible with the given data within noise
bounds~\cite{Brown2003}.  (See Appendix for details.)
After fitting to
the experimental results shown in Figure~6A, the resulting ensemble of
parameter sets can be used to make ensemble predictions for novel
experimental conditions; for example, we can estimate the possible
range of concentrations of any chemical species at any time point
given initial conditions.

Using the ensemble parameters fitted to the training set (Figure~6A), the simulation results for Figure~5 were generated and compared with the experimental results (Figure~6B).  The kinetic model predicted that the switch would either approach the low steady-state in which all A were bound to dI and the rA concentrations converged to 0.5~$\mu$M or the high steady-state in which all A were free from dI and the rA concentrations kept increasing.
The experimentally measured fluorescence signal from A and sampled rA concentrations were close to the prediction (Figure~6B), despite the fact that the training data set contained information on rA concentrations at a single time point only.  However, in the experiment, the rA concentrations for the ON state did not increase indefinitely as the model predicted, but seemed to reach a high steady-state. We shall address this discrepancy between the model and experiment later.

\begin{figure*}[tbh]
\centerline{\epsfig{figure=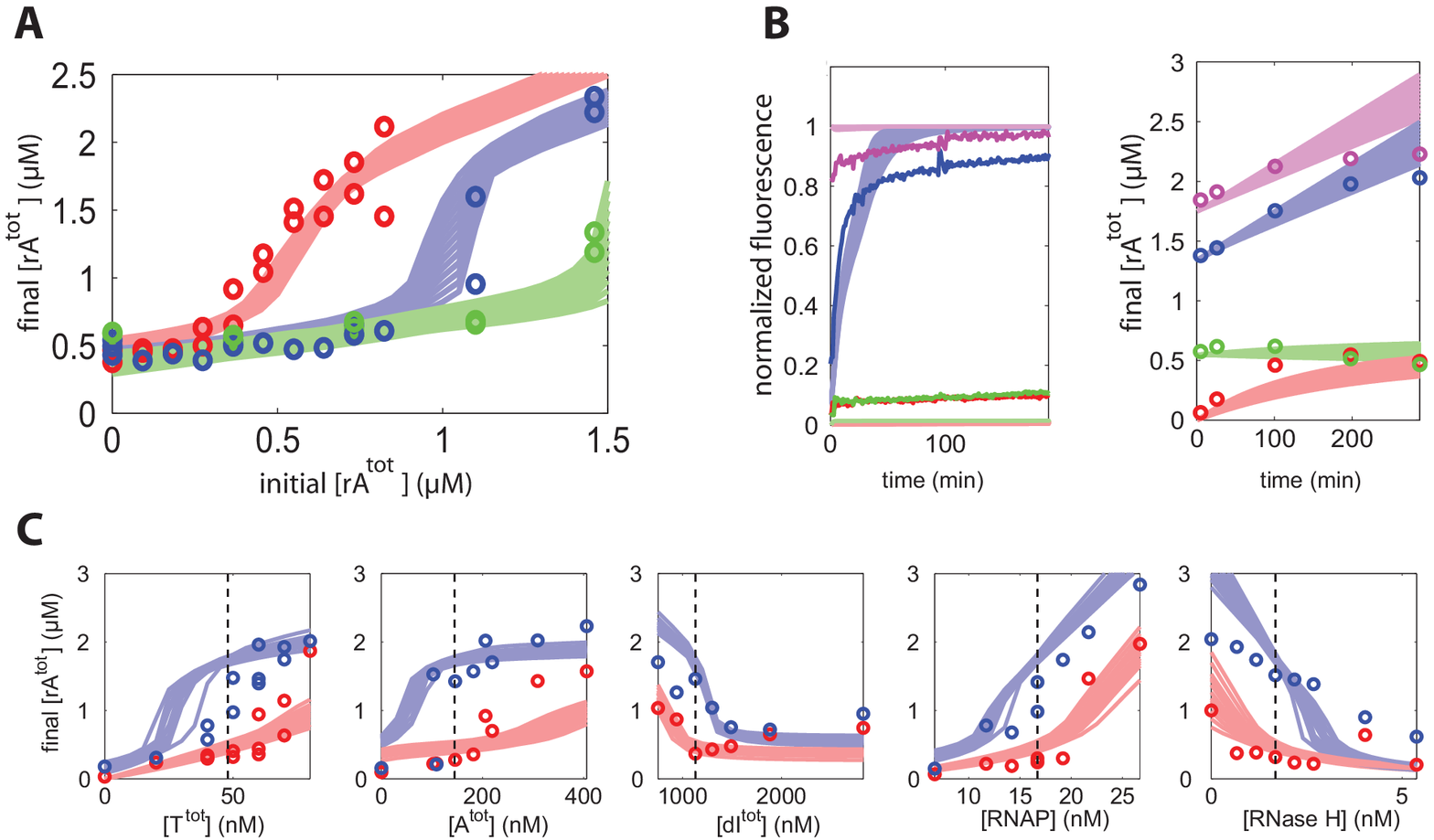, width = 0.95 \textwidth}}
\caption{\textsf{Quantitative measurements and model prediction of the self-activating switch dynamics. The circles within each figure and the lines with darker colors in (B, left) are from experimental measurements.  Lines with lighter shades are generated from 95\% confidence interval of model parameters fitted to the experimental results shown in (A).  Experimental results in (B) and (C) are used as test cases of model prediction.
(\textbf{A}) The training data set used for parameter fitting.   The circles are gel measurements of final [rA${\rm ^{tot}}$] at 210 min.
All reactions contained 48 nM of [T${\rm ^{tot}}$], 145 nM of [A${\rm ^{tot}}$], 16.7 nM of [RNAP], and 1.68 nM of [RNaseH].
The concentration of dI was varied as follows:
[dI${\rm ^{tot}}$] = 1.13~$\mu$M (red), 1.45~$\mu$M (blue), or 1.87~$\mu$M (green). Note that the thresholds of switching between low and high states were apparently shifted as the total concentration of dI was increased.
(\textbf{B})
Experimental and predicted time courses of switch states (monitored as fluorescence changes) and rA concentrations. The experimental data is identical to Figure~5. The experimental conditions were similar to those colored blue in (A): only dI was 50~nM higher for (B).
 (\textbf{C})
Experimental and predicted parameter dependence of self-activation switch behavior.
The default set of experimental conditions were identical to those colored red in (A): [T${\rm ^{tot}}$]~=~48~nM, [A${\rm ^{tot}}$]~=~145~nM, [dI${\rm ^{tot}}$]~=~1.13~$\mu$M, [RNAP]~=~16.7~nM, and [RNaseH]~=~1.68~nM (marked as black dashed lines in each subplot). For each experimental measurement, one parameter was varied at a time and the final concentration of rA after 210 min was recorded with initial rA concentration of 0~nM (red) or 730~nM (blue), respectively.
}}
\end{figure*}

To further test the sensitivity of system behavior to parameter variation, we systematically varied the template ([T${\rm ^{tot}}$]), activator ([A${\rm ^{tot}}$]), inhibitor ([dI${\rm ^{tot}}$]), and enzyme concentrations ([RNAP] and [RNaseH]) (Figure~6C). For each parameter choices, we initiated the transcription and degradation reaction with either a low amount of rA (0~nM) or a high amount of rA (730~nM).
If the final difference of RNA activators is greater than the initial difference of 730~nM, it would imply that these two initial conditions lead to different switch states.  However, since the switch states were not monitored by fluorescence, these results were not a direct readout of bistable switch response.  On the other hand, if both initial conditions lead to low steady-states, the switch could be in the monostable regime, but it also could be in a bistable regime that requires higher initial rA inputs to reach high steady-state. Yet another possibility is when both initial conditions lead to high steady-states such that the differences of final rA concentrations would be close to 730~nM.  Thus, for experimental results, we marked out the parameter ranges where the differences of final rA concentrations were greater than 800~nM as indications of bistable switch responses.
The model predicted that the self-activating switch showed a bistable response for a limited range of parameter variations near the default values. For simulation results, we map the parameter range as bistable, if the final switch states were different for the 95\% of ensemble parameter values. 
From the simulation results, the self-activating switch reached the OFF state under conditions with low T, low A, low RNAP, high dI, or high RNase~H, irrespective of initial conditions.  Conversely, the self-activating switch reached the ON state under conditions with high T, high A, high RNAP, low dI, or low RNase~H, irrespective of initial conditions.
The experimental results were in good agreement with the model prediction on the range of bistability for each parameter. 
Not surprisingly, the model predictions showed increased uncertainty in the transition regions for both high and low initial rA concentrations.

\vspace{.1in}
\noindent{\textbf{Discrepancies in modeling}}
\vspace{.1in}

Given its simplicity, our model predicted the dynamics of the switch
surprisingly well.  We only included the main hybridization and enzyme
reactions that constituted our design intentions, excluding the
unintentional `interfering,' `recovering,' and `recapturing'
reactions as well as any other undesired crosstalk between DNA or RNA
species.  Notably, the training data used for fitting (Figure~6A) only
gives information about how the final [rA${\rm ^{tot}}$] depends on the
initial [rA${\rm ^{tot}}$] and [dI${\rm ^{tot}}$], and thus provides very weak
kinetic constraints on the model.  Unsurprisingly, a wide range of
parameter sets could fit the data within reasonable noise bounds, as
is common when fitting biochemical system models to limited
data~\cite{gutenkunst2007universally}.  Nonetheless, qualitative (and
sometimes quantitative) predictions were made accurately for a variety
of novel experimental conditions (the test data, Figure 6B,C),
including the time evolution of the RNA signal and switch state.  This
suggests that the intended reaction pathways, used to construct the
model, are dominant ones in the system dynamics.

However, some predictions were qualitatively wrong for all parameter
sets in the {\it a posteriori} ensemble.  These prediction errors may
point to omissions in the model.  As an example, for the time
evolution of the RNA signal (Figure 6B, right), the model predicted that
the upper two reactions would reach a state in which [${\rm rA^{tot}}$]
keeps increasing, while in the experiment, these two reactions seemed
to converge to a steady-state.  A possible explanation is that
incomplete degradation by RNase~H, as observed in previous
work~\cite{Kim2006}, yields short fragments of rA that accumulate during the reaction: at high concentrations of these short products, product rebinding to the transcription initiation complex of RNAP may decrease transcription rates~\cite{Kuzmine2001}.

As another example, the model predicts a stronger sensitivity to
RNase~H concentration than was observed (Figure 6C, rightmost panel).
Incomplete degradation fragments may also provide an explanation for
this discrepancy.  These shorter oligonucleotides could also bind to
dI and thus compete for the degradation by RNase~H, decreasing its effectiveness.
Other notable discrepancies are that the model predicts that wider
ranges of [${\rm T^{tot}}$] and [${\rm A^{tot}}$] will sustain
bistability than was experimentally observed.  Another
oversimplification of the model is that the inhibition and release
reactions are expected to be reversible, despite being
thermodynamically favored in the forward direction as described; they
are instances of so-called toehold exchange, as there is also a
(weaker) toehold in the reverse direction as
well~\cite{zhang2009exchange}.

\vspace{.3in}
\noindent{\Large \bf Conclusions}
\vspace{.1in}

In this work we have designed and synthesized an activating cascade
for {\it in vitro} transcriptional circuits and demonstrated its use
as a single-switch bistable system.  In comparison to the previously
constructed bistable circuit comprised of two mutually inhibiting
switches~\cite{Kim2006}, our single-switch system is simpler (four
strands vs six) but more delicate to design.  DNA hybridization
reactions can no longer be made arbitrarily strongly favored and
irreversible; the `inhibition' and `release' reactions must therefore
be designed carefully.  Further, the multistep activating cascade
makes use of both `sense' and `antisense' sequences, making it
difficult to eliminate spurious reactions such as `interfering',
`recovering', and `recapturing'.  Fortunately, the working solution
demonstrated here ought to be easily generalized to other sequences,
enabling the modular construction of {\it in vitro} transcriptional
circuits containing both activating and inhibiting switches.

Fulfilling the promise of predictable system design, when configured
to self-activate by using the same sequence elements for both input
and output domains, the expected bistable behavior was observed.
Somewhat more surprisingly, when a simple network model was trained on
limited data, system sensitivities to DNA and enzyme concentrations
were qualitatively and sometimes quantitatively predicted.  These
studies also suggest that the addition of a single-stranded RNA
exonuclease, such as RNase~R, would help clean up incomplete
degradation fragments, improve performance, and yield a more
predictable system.

While in principle arbitrary behaviors can be obtained using networks
of inhibiting switches (for example, activation can be obtained by two
inhibiting switches in series), the direct implementation of
activation using a single switch provides some advantages.  As
discussed before, it yields simpler circuits in terms of the number of
DNA strands required.  Furthermore, the hybridization reactions are
faster and more energy efficient than the transcription reactions they
replace.  This observation raises a question for the future
construction of larger {\it in vitro} circuits: how much of the
desired logic should be implemented by transcriptional circuitry~\cite{Kim2004,Kim2006}
and how much should be implemented as strand displacement
circuitry~\cite{Seelig2006logic,Zhang2007entropy,Soloveichik2010crns}?

\vspace{.3in}
\noindent{\Large \bf Materials and Methods}

\vspace{.1in}
\noindent
\noindent{\bf DNA oligonucleotides and enzymes}
\vspace{.1in}

{ The sequence of all DNA molecules and expected RNA transcript sequences were chosen to minimize the occurrence of alternative secondary structures, checked by the Vienna group's DNA  and RNA folding program~\cite{flamm00}.  All DNA oligonucleotides were purchased from Integrated DNA Technologies (USA). In addition to unlabeled A and dI, A labeled with Cy5 at the 3' end and dI labeled with IowaBlack-RQ at the 5' end were purchased.  The T7 RNA polymerase (enzyme mix), transcription buffer, and NTP were purchased as part of the T7 Megashortscript kit (Ambion, Austin, Texas, USA; \#1354).  RNase~H (Ambion; \#2293) was purchased.  RNase~R was a gift from Dr.~Murray P.~Deutscher at University of Miami, school of Medicine. Note that according to the manufacturer's patent (\# 5,256,555), the enzyme mix of T7 Megashortscript kit contains pyrophosphatase to extend the lifetime of the transcription reaction; since pyrophosphatase is involved in regulating the power supply for our transcriptional circuits and is not directly involved in the dynamics, we neglect this enzyme in our models and do not call it an ``essential enzyme'' for the circuit dynamics.}

\vspace{.1in}
\noindent
{\bf Transcription}
\vspace{.1in}

{ DNA templates (T-nt and T-t strands) were annealed with 10\% (v/v) 10x transcription buffer from 90$^\circ$C to 20$^\circ$C over 1 hour at 10 times the final concentration used.  To the annealed templates, DNA activator and inhibitor, A and dI, were added from high concentration stocks (50~$\mu$M), 7.5~mM each NTP, and 8\% (v/v) 10x transcription buffer were added.   After adding all ssDNA strands and RNA signals, enzymes (RNAP, RNase~H, and RNase~R) were added and mixed. Transcription reactions for spectrofluorometer experiments were prepared as a total volume of  60~$\mu$L.  Transcription reactions for gel studies were prepared as a total volume of 10~$\mu$L and were stopped by denaturing dye (80\%  formamide, 10~mM EDTA, 0.01~g XCFF). For the purification of RNA signal rA for the experiments and for gel controls,
the full length template side strand (the complement of T-nt
rather than T-t) was used to prepare a fully duplex DNA template
for rA. The transcription reaction was prepared as a total volume of
60~$\mu$L with 0.2~$\mu$M fully duplex DNA templates.  The
transcription condition was the same as above except that no A or dI were added, 20\% (v/v)
RNAP was used, and RNases were omitted. After a 6 hour
incubation at 37$^\circ$C, the reaction mixture was treated with
2.5~$\mu$L DNase~I for 30 min to remove DNA template, and
stopped by the addition of denaturing dye.  The reaction mixture was
run on 8\% denaturing gel, RNA bands were excised and
eluted from gel by the crush-and-soak method, ethanol precipitated, and resuspended in water.}

\vspace{.1in}
\noindent
{\bf Data acquisition}
\vspace{.1in}

{ For spectrofluorometer experiments, excitation and emission for Cy5-labeled A were at 648~nm and 665~nm.
The fluorescence was recorded every minute using a SPEX Fluorolog-3 (Jobin Yvon, Edison, New Jersey, United States) and normalized against maximum fluorescence (measured in the presence of Cy5-labeled A prior to the addition of quencher-labeled dI) and minimum fluorescence (measured after the addition of excess quencher-labeled dI) accounting for volume increase due to the addition of rA and enzymes.  Denaturing polyacrylamide gels (8\% 19:1 acrylamide:bis and 7~M urea in TBE buffer) were allowed to run for 50~min with 10~V/cm at 65$^\circ$C in TBE buffer (100~mM Tris, 90~mM Boric Acid, 1~mM EDTA). The 10-base DNA ladder (Invitrogen, Carlsbad, California, United States; \#10821-015) was used in the control lane.
The non-denaturing gels (10\% 19:1 acrylamide:bis in TAE buffer) were allowed to run for 100 min with 13V/cm at 35$^\circ$C in TAE buffer containing 12.5~mM Mg$^{2+}$ (40~mM Tris-Acetate, 1~mM EDTA, 12.5~mM  Mg-Acetate, pH 8.3). The gels were stained with SYBR gold (Molecular Probes, Eugene, Oregon, United States; \#S-11494) and the gel data was quantitated using the Molecular imager FX (Biorad, Hercules, California, United States). The total concentrations of RNA product rA in the denaturing gel were measured with respect to 0.5, 1, and 1.5~$\mu$M purified rA in control lanes. }

\vspace{.1in}
\noindent
{\bf Model simulation}
\vspace{.1in}

{The kinetic simulations and parameter fittings were implemented in MATLAB. Differential equations were solved using the {\em ode23s} routine.
The parameter sets used in Figure~6 were derived from the Metropolis' minimization of sum squared differences between the experimental results in Figure~6A and the simulation results of kinetic model described in Appendix. We identify model parameters, $\vec{\theta}$, as a vector of logs of rate constants; the kinetic model has 11 parameters ($k_{TA}$, $k_{AI}$, $k_{rAI}$, $k_{TAI}$, $k_{AIrA}$, $k_{M,ON}$, $k_{M, OFF}$, $k_{cat, ON}$, $k_{cat, OFF}$, $k_{M,H}$, and $k_{cat, H}$). Each parameter has pre-specified minimum and maximum values (see Appendix).  The cost function is defined as  \[ E = \sum_{n=1}^N ([rA]_{exp}^{final,n} - [rA]_{sim}^{final,n})^2, \] where the concentration of rA is measured at 210 min in $\mu$M scale both for the experiment and simulation.  The training set (Figure~6A) contained fifty experimentally measured [rA] values ($N$ = 50).  Starting from random initial values, for each iteration, a random 11x1 vector of -1, 0, or +1 (with equal probabilities) was generated to decide whether to decrease, remain, or increase each parameter in the next step. The step size for each parameter was 1/100th of the range for that parameter in log scale.
We accepted or rejected the next step based on Metropolis criteria --- i.e., the acceptance probability was $p$ = 1 if $\Delta E < 0$ and $p$ = $e^{-\Delta E/T}$ if $\Delta E > 0$ with T = 2$\sigma^2$ = 0.125 $\mu$M$^2$. We used three sampling trajectories with 620,000 iteration steps starting from random parameter values, discarding the first 20\% of iterations and sampling one parameter set per 30,000 iterations. This collection of parameter sets ($M$ = 51) was used for simulation results in Figure~6; they show the 95\% confidence interval for the parameter values --- i.e., the four (out of 51 parameter sets) with largest $E$ are omitted. See Appendix for details.}

\vspace{.1in}
\noindent
{\bf DNA sequences}
\vspace{.1in}

\noindent
T-nt (106 mer), {\tiny
5'CATTAGTGTCGTTCGTTCACAGTAATACGACT\-CA\-CT\-ATAGGGAGAAACAAAGAACGAACGACACTAATGAACTACTACTAC\-A\-CA\-CT\-AATACTGACAAAGTCAGAAA-3'.}

\noindent
T-t (79 mer), {\tiny
5'TTTCTGACTTTGTCAGTATTAGTGTGTAGTAG\-TAGT\-TCATTAGTGTCGTTCGTTCTTTGTTTCTCCCTATAGTGAGTCG-3'.}

\noindent
A (36 mer), {\tiny   	
5'TATTACTGTGAACGAACGACACTAATGAACTACTAC-3'.}

\noindent
dI (38 mer), {\tiny
5'-GTGTGTAGTAGTAGTTCATTAGTGTCGTTCGTTC\-AC\-AG-3'.}

\noindent
rA (67 mer), {\tiny
5'GGGAGAAACAAAGAACGAACGACACUAAUGAAC\-UA\-CUACUACACACUAAUACUGACAAAGUCAGAAA 3'.}
\\

Note that the RNA activator rA sequence is identical to I1 in~\cite{Kim2006}, and therefore, T-t sequence is also identical to T12-t in~\cite{Kim2006}.
{\scriptsize
\bibliographystyle{plain}
\bibliography{journalshort,SelfActivatingSwitch}

\begin{thebibliography}{10}

\bibitem{Andr2006}
E.~Andrianantoandro, S.~Basu, D.~K. Karig, and R~Weiss.
\newblock Synthetic biology: new engineering rules for an emerging discipline.
\newblock {\em Mol. Syst. Biol.}, 2:2006.0028, 2006.

\bibitem{Atkinson2003}
M.~R. Atkinson, M.~A. Savageau, J.~T. Myers, and A.~J. Ninfa.
\newblock Development of genetic circuitry exhibiting toggle switch or
  oscillatory behavior in {{\it Escherichia coli}}.
\newblock {\em Cell}, 113:597--607, 2003.

\bibitem{Bayer2005}
Travis~S Bayer and Christina~D Smolke.
\newblock Programmable ligand-controlled riboregulators of eukaryotic gene
  expression.
\newblock {\em Nature Biotech.}, 23(3):337--343, 2005.

\bibitem{Becskei2001}
A.~Becskei, B~Seraphin, and L~Serrano.
\newblock Positive feedback in eukaryotic gene networks: cell differentiation
  by graded to binary response conversion.
\newblock {\em EMBO J}, 20:2528--2535, 2001.

\bibitem{Benner2005}
Steven~A. Benner and A.~Michael Sismour.
\newblock Synthetic biology.
\newblock {\em Nature Rev. Genet.}, 6:533--543, 2005.

\bibitem{Brown2003}
Kevin~S Brown and James~P Sethna.
\newblock Statistical mechanical approaches to models with many poorly known
  parameters.
\newblock {\em Phys. Rev. E}, 68:021904, 2003.

\bibitem{Elowitz00}
Michael~B. Elowitz and Stanislas Leibler.
\newblock A synthetic oscillatory network of transcriptional regulators.
\newblock {\em Nature}, 403:335--338, 2000.

\bibitem{Endy2005}
Drew Endy.
\newblock Foundations for engineering biology.
\newblock {\em Nature}, 438:449--453, 2005.

\bibitem{flamm00}
C.~Flamm, W.~Fontana, I.~Hofacker, and P.~Schuster.
\newblock {RNA} folding at elementary step resolution.
\newblock {\em RNA}, 6:325--338, 2000.

\bibitem{Gao2006}
Yang Gao, Lauren~K. Wolf, and Rosina~M. Georgiadis.
\newblock Secondary structure effects on {DNA} hybridization kinetics: a
  solution versus surface comparison.
\newblock {\em Nucleic Acids Res}, 34:3370--3377, 2006.

\bibitem{gutenkunst2007universally}
R.N. Gutenkunst, J.J. Waterfall, F.P. Casey, K.S. Brown, C.R. Myers, and J.P.
  Sethna.
\newblock {Universally sloppy parameter sensitivities in systems biology
  models}.
\newblock {\em PLoS Comput Biol}, 3(10):1871--1878, 2007.

\bibitem{Hasty2001}
J.~Hasty, D.~McMillen, and J.~J. Collins.
\newblock Engineering gene circuits.
\newblock {\em Nature}, 420:224--230, 2001.

\bibitem{Hooshangi2005}
Sara Hooshangi, Stephan Thiberge, and Ron Weiss.
\newblock Ultrasensitivity and noise propagation in a synthetic transcriptional
  cascade.
\newblock {\em Proc. Natl. Acad. Sci. USA}, 102(10):3581--3586, 2005.

\bibitem{Isaacs2006}
Farren~J Isaacs, Daniel~J Dwyer, and James~J Collins.
\newblock {RNA} synthetic biology.
\newblock {\em Nature Biotech.}, 24:545--554, 2006.

\bibitem{Isaacs2004}
Farren~J Isaacs, Daniel~J Dwyer, Chunming Ding, Dmitri~D Pervouchine, Charles~R
  Cantor, and James~J Collins.
\newblock Engineered riboregulators enable post-transcriptional control of gene
  expression.
\newblock {\em Nature Biotech.}, 22(7):841--847, 2004.

\bibitem{Isaacs2003}
Farren~J Isaacs, Jeff Hasty, Charles~R Cantor, and James~J Collins.
\newblock Prediction and measurement of an autoregulatory genetic module.
\newblock {\em Proc. Natl. Acad. Sci. USA}, 100:7714--7719, 2003.

\bibitem{jewett2008integrated}
M.C. Jewett, K.A. Calhoun, A.~Voloshin, J.J. Wuu, and J.R. Swartz.
\newblock {An integrated cell-free metabolic platform for protein production
  and synthetic biology}.
\newblock {\em Molecular Systems Biology}, 4(1):220, 2008.

\bibitem{Jia1997}
Yiping Jia and Smita~S. Patel.
\newblock Kinetic mechanism of transcription initiaion by bacteriophage {T7
  RNA} polymerase.
\newblock {\em Biochemistry}, 36:4223--4232, 1997.

\bibitem{Kim2004}
Jongmin Kim, John~J. Hopfield, and Erik Winfree.
\newblock Neural network computation by {\it in vitro} transcriptional
  circuits.
\newblock In {\em Advances in Neural Information Processing Systems (NIPS)},
  volume~17, pages 681--688, 2004.

\bibitem{Kim2006}
Jongmin Kim, Kristin~S White, and Erik Winfree.
\newblock Construction of an {\it in vitro} bistable circuit from synthetic
  transcriptional switch.
\newblock {\em Mol. Syst. Biol.}, 2:68, 2006.

\bibitem{Kuzmine2001}
Iaroslav Kuzmine and Craig~T Martin.
\newblock Pre-steady-state kinetics of initiation of transcription by {T7 RNA}
  polymerase: {A} new kinetic model.
\newblock {\em J. Mol. Biol.}, 305:559--566, 2001.

\bibitem{Marras2002}
Salvatore A.~E. Marras, Fred~Russell Kramer, and Sanjay Tyagi.
\newblock Efficiencies of fluorescence resonance energy transfer and
  contact-mediated quenching in oligonucleotide probes.
\newblock {\em Nucleic Acids Res.}, 30(21):e122, 2002.

\bibitem{Martin1987}
Craig~T. Martin and Joseph~E. Coleman.
\newblock Kinetic analysis of {T7} {RNA} polymerase-promoter interactions with
  small synthetic promoters.
\newblock {\em Biochemistry}, 26:2690--2696, 1987.

\bibitem{McGinness2002}
Kathleen~E McGinness and Gerald~F Joyce.
\newblock Substitution of ribonucleotides in the {T7 RNA} polymerase promoter
  element.
\newblock {\em J Biol Chem}, 277:2987--2991, 2002.

\bibitem{Mills1999}
A.~P. {Mills Jr.}, Bernard Yurke, and Philip~M Platzman.
\newblock Article for analog vector algebra computation.
\newblock {\em Biosystems}, 52:175--180, 1999.

\bibitem{Noireaux2003}
Vincent Noireaux, Roy Bar-{Z}iv, and Albert Libchaber.
\newblock Principles of cell-free genetic circuit assembly.
\newblock {\em Proc. Natl. Acad. Sci. USA}, 100(22):12,672--12,677, 2003.

\bibitem{Seelig2006logic}
G.~Seelig, D.~Soloveichik, D.~Y. Zhang, and E.~Winfree.
\newblock Enzyme-free nucleic acid logic circuits.
\newblock {\em Science}, 314(5805):1585--1588, 2006.

\bibitem{Seelig2005}
G.~Seelig, B.~Yurke, and E.~Winfree.
\newblock {DNA} hybridization catalyst and catalyst circuit.
\newblock {\em DNA comp}, 3384:329--343, 2005.

\bibitem{Seeman82}
Nadrian~C. Seeman.
\newblock Nucleic-acid junctions and lattices.
\newblock {\em J. Theor. Biol.}, 99:237--247, 1982.

\bibitem{Shimizu2001}
Yoshiru Shimizu, Akio Inoue, Yukihide Tomari, Tsutomu Suzuki, Takashi Yokogawa,
  Kazuya Nishikawa, and Takuya Ueda.
\newblock Cell-free translation reconstituted with purified components.
\newblock {\em Nature Biotech.}, 19:751--755, 2001.

\bibitem{Soloveichik2010crns}
David Soloveichik, Georg Seelig, and Erik Winfree.
\newblock {DNA} as a universal substrate for chemical kinetics.
\newblock {\em Proc. Natl. Acad. Sci. USA}, 107(12):5393--5398, 2010.

\bibitem{Stricker2008}
Jesse Stricker, Scott Cookson, Matthew~R. Bennett, William~H. Mather, Lev~S.
  Tsimring, and Jeff Hasty.
\newblock A fast, robust and tunable synthetic gene oscillator.
\newblock {\em Nature}, 456:516--519, 2008.

\bibitem{Triana-alonso1995}
Francisco~J. Triana-Alonso, Marylena Dabrowski, Jorg Wadzack, and Knud~H.
  Nierhaus.
\newblock Self-coded 3'-extension of run-off transcripts produces abberant
  products during {{\it in vitro}} transcription with {T7} {RNA} polymerase.
\newblock {\em J. Biol. Chem.}, 270:6298--6307, 1995.

\bibitem{Win2007}
Maung~Nyan Win and Christina~D Smolke.
\newblock A modular and extensible {RNA}-based gene-regulatory platform for
  engineering cellular function.
\newblock {\em Proc. Natl. Acad. Sci. USA}, 104(36):14283--14288, 2007.

\bibitem{Zhang2007entropy}
D.~Y. Zhang, A.~J. Turberfield, B.~Yurke, and E.~Winfree.
\newblock Engineering entropy-driven reactions and networks catalyzed by {DNA}.
\newblock {\em Science}, 318(5853):1121--1125, 2007.

\bibitem{zhang2009exchange}
D.~Y. Zhang and E.~Winfree.
\newblock Control of {DNA} strand displacement kinetics using toehold exchange.
\newblock {\em J Am Chem Soc}, 131(47):17303--17314, 2009.

\end{thebibliography}
}

\newpage
%
\onecolumn
\newpage
\noindent
{\large{\bf Appendix}} \\
\\
{\large \em A simple kinetic model for Self-activating switch}\\

To mathematically explore the behavior of the self-activating switch, we developed a model based on the DNA and RNA hybridization reactions, and enzyme reactions. Note that for the self-activating switch, the input RNA signal is identical to the output RNA signal, thus both are labeled rA. We constructed a mathematical model without RNase R, because RNase~R was not used for the results presented in Figures~5 and 6 where the model is used. 
To simplify the model, we also assumed that, although an `interfering' reaction exists that results in the T$\cdot$rA complex, the `recovering' and `recapturing' reactions involving A and dI are fast enough such that the concentration of the T$\cdot$rA complex is negligible (see Figure~3). The hybridization and strand displacement reactions and enzyme reactions are summarized below. \\\\

\noindent
{\em DNA or RNA hybridizations}
$$ \begin{array}{rclr}
{\rm T} + {\rm A} & \stackrel{k_{TA}}{\rightarrow}  & {\rm T}\cdot{\rm A} & {\rm (Activation)}\\
{\rm A} + {\rm dI} & \stackrel{k_{AI}}{\rightarrow} & {\rm A}\cdot{\rm dI}  &{\rm (Annihilation)}\\
{\rm rA} + {\rm dI} & \stackrel{k_{rAI}}{\rightarrow} & {\rm rA}\cdot{\rm dI}  &{\rm (Annihilation)}\\
{\rm dI}+{\rm T}\cdot{\rm A}& \stackrel{k_{TAI}}{\rightarrow} & {\rm T} + {\rm A}\cdot{\rm dI}   &{\rm (Inhibition)}\\
{\rm rA} + {\rm A}\cdot{\rm dI} & \stackrel{k_{AIrA}}{\rightarrow} & {\rm rA}\cdot{\rm dI} + {\rm A}  &{\rm (Release)}\\
\end{array}$$

\noindent
{\em Michaelis--Menten Enzyme reactions}
$$   \begin{array}{rclr}
{\rm T}\cdot{\rm A}   &\stackrel{Prod_{ON}}{\rightarrow} & {\rm T}\cdot{\rm A} + {\rm rA} & {\rm (transcription, fast)}  \\
{\rm T} & \stackrel{Prod_{OFF}}{\rightarrow}& {\rm T} + {\rm rA} & {\rm (transcription, slow)} \\
{\rm rA}\cdot{\rm dI} & \stackrel{Deg_{H}}{\rightarrow}  &{\rm dI}   & {\rm (degradation)} \\
\end{array} $$

We do not consider side-reactions or incomplete production and degradation products. We further simplified enzymatic equations from the full Michaelis-Menten  equations to an approximation using the pseudo-steady-state assumption of enzyme-substrate complex, which is reasonably accurate when enzyme concentrations are low compared to substrate concentrations.

$$\begin{array}{rcccl}
{\rm RNAP} + {\rm T}\cdot{\rm A} &  \overset{k_+}{\underset{k_{-,ON}}{\rightleftharpoons}} & {\rm RNAP}\cdot{\rm T}\cdot{\rm A} & \stackrel{k_{cat,ON}}{\rightarrow} &  {\rm RNAP} + {\rm T}\cdot{\rm A} + {\rm rA} \\
{\rm RNAP} + {\rm T} & \overset{k_+}{\underset{k_{-,OFF}}{\rightleftharpoons}} & {\rm RNAP}\cdot{\rm T} &  \stackrel{k_{cat,OFF}}{\rightarrow} & {\rm RNAP} + {\rm T} + {\rm rA} \\
{\rm RNase H} + {\rm rA}\cdot{\rm dI}  & \overset{k_{+,H}}{\underset{k_{-,H}}{\rightleftharpoons}}  & {\rm RNase H}\cdot{\rm rA}\cdot{\rm dI} & \stackrel{k_{cat,H}}{\rightarrow} & {\rm RNase H} + {\rm dI} \\
\end{array}$$

\noindent
We express the available enzyme concentrations using the pseudo-steady-state assumption as follows:
\begin{equation*}
[{\rm RNAP}]=\frac{[{\rm RNAP^{tot}}]}{1+\sum{\frac{[{\rm T}\cdot{\rm A}]}{K_{M,ON}}+\sum{\frac{[{\rm T}]}{K_{M,OFF}}}}}, \qquad [{\rm RNase H}]=\frac{[{\rm RNase H^{tot}}]}{1+\frac{[{\rm rA}\cdot {\rm dI}]}{K_{M,H}}},
\end{equation*}
where the Michaelis constants are calculated as $K_M=\frac{k_- + k_{cat}}{k_+}$ to determine the affinity of substrates to the enzymes.
The effective rate constants for enzymes are as follows:

\begin{equation*}
{Prod_{ON}}=\frac{k_{cat, ON}}{K_{M, ON}}{\rm [RNAP]} , \quad
{Prod_{OFF}}=\frac{k_{cat, OFF}}{K_{M, OFF}}{\rm [RNAP]}, \quad
{Deg_{H}}=\frac{k_{cat, H}}{K_{M, H}}{\rm [RNaseH]}. \quad
\end{equation*}

Note that this approximation procedure was also used in~\cite{Kim2006}.
Taken together, a set of four ordinary differential equations describes the dynamics of the self-activating switch as follows.

\begin{eqnarray*}
\frac{d[{{\rm T}}]}{dt} &=& - k_{TA}[{\rm T}][{\rm A}] + k_{TAI}[{\rm T}\cdot{\rm A}][{\rm dI}] \\
\frac{d[{{\rm A}}]}{dt} &=& - k_{AI}[{\rm A}][{\rm dI}] - k_{TA}[{\rm T}][{\rm A}] + k_{AIrA}[{\rm A}\cdot{\rm dI}][{\rm rA}] \\
\frac{d[{{\rm dI}}]}{dt} &=& - k_{AI}[{\rm A}][{\rm dI}] - k_{rAI}[{\rm rA}][{\rm dI}] - k_{TAI}[{\rm T}\cdot{\rm A}][{\rm dI}] + Deg_{H}[{\rm rA}\cdot{\rm dI}] \\
\frac{d[{{\rm rA}}]}{dt} &=& - k_{rAI}[{\rm rA}][{\rm dI}] - k_{AIrA}[{\rm A}\cdot{\rm dI}][{\rm rA}] + Prod_{ON}[{\rm T} \cdot{\rm A}] + Prod_{OFF}[{\rm T}]
\end{eqnarray*}

\noindent
The remaining variables, [T$\cdot$A], [A$\cdot$dI], and [rA$\cdot$dI] are calculated from the conservation relation as  [T${\rm ^{tot}}$], [A${\rm ^{tot}}$], and  [dI${\rm ^{tot}}$]  remain constants throughout the reaction. \\

%
%
%

\renewcommand{\thefigure}{S\arabic{figure}}
\setcounter{figure}{0}

\noindent
{\large \em Sampling parameter space with Metropolis criteria}\\

Using the mathematical model described above, we chose to explore the parameter space in the vicinity of the best fit to data based on the Metropolis sampling method, as outlined in~\cite{Brown2003}. The experimental data shown in Figure~6A is used as the training set.  Other experimental data are reserved as the test set.  The cost function is calculated as
\[ E = \sum_{n=1}^N ([rA]_{exp}^{final,n} - [rA]_{sim}^{final,n})^2,
\] where the concentration of rA is measured at 210 min in $\mu$M scale both for the experiment and simulation.  The training set contained fifty experimentally measured [rA] values ($N$ = 50).  To estimate errors in experimental measurements, most experimental conditions were measured in duplicate.  For certain experimental conditions --- e.g.~in the transition region of blue curve in Figure~6A --- the duplicate measurements differed as much as 0.65~$\mu$M.  We chose $\sigma$ = 0.25~$\mu$M as a reasonable estimate of the standard deviation of repeated experimental measurements.

The kinetic model has a total of 11 parameters: $k_{TA}$, $k_{AI}$, $k_{rAI}$, $k_{TAI}$, $k_{AIrA}$, $k_{M,ON}$, $k_{M, OFF}$, $k_{cat, ON}$, $k_{cat, OFF}$, $k_{M,H}$, and $k_{cat, H}$. We set the lower and upper bounds for these parameters as (in log$_{10}$ scale) [4.5, 4.5, 4.5, 4.5, 4.5, -8, -7, -2, -3, -7, -2] and [6.0, 6.0, 6.0, 6.0, 6.0, -6, -5, -1, -1, -5, 0], respectively.  In order to minimize the effect of these widely separated scales and avoid numerical issues, we deal with the logarithms of the parameters for all our calculations. Henceforth, we identify model parameters, $\vec{\theta}$, as a vector of logs of rate constants. We initiated the parameter set $\vec{\theta}$ as a random 11x1 vector within these bounds. For each iteration, we generated a random 11x1 vector of -1, 0, or +1 (with equal probabilities) to decide whether to decrease, remain, or increase each parameter in the next step. The step size for each parameter was set as 1/100th of the range for that parameter in a log scale. If the simultaneous updates of 11 parameters resulted in lower cost function (i.e.~$\Delta E < 0$), the updates are accepted. On the other hand, if the simultaneous updates of parameters resulted in higher cost function (i.e.~$\Delta E > 0$), the updates are accepted with a probability of $p = e^{-\Delta E/T}$ following Metropolis criteria. We accepted or rejected the parameter updates based on the Metropolis criteria with T = 0.125 (= 2$\sigma^2$).

Following these sampling procedures, 620,000 iterations were performed for trajectories 1 through 3 and 470,000 iterations were performed for trajectory 4 starting from random initial conditions.
To assess whether there are multiple attractor basins for parameter values, we plotted parameter values for the whole trajectories as histograms (Figure~S1) with the initial values marked by red circles. The parameter distributions for all 11 parameters achieved similar ranges and shapes irrespective of their initial values.  Therefore, we conclude that the Monte Carlo sampling with Metropolis criteria has converged for these trajectories.

\newpage
\begin{figure*}[h!!]
\centerline{\epsfig{figure=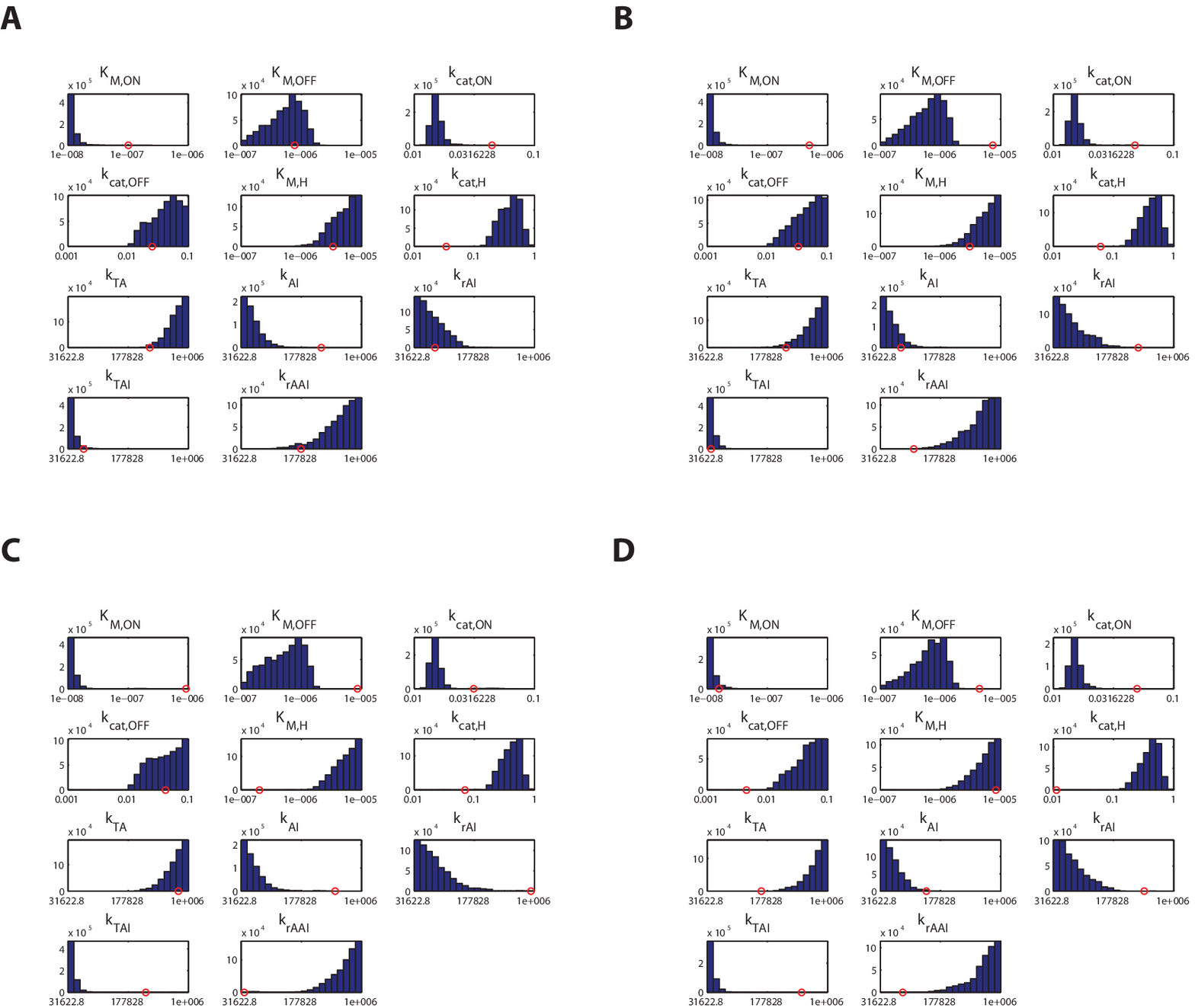,width=.95\textwidth}}
\caption{\textsf{Histograms for parameter-sampling trajectories. ({\textbf{A--D}}) The log-scale parameter values ($\vec{\theta}$) for the whole sampling trajectories are plotted as histograms. The x-axes are in log scale with the respective minimum and maximum values corresponding to the range of each parameter as stated earlier. Following Metropolis criteria, 620,000 iterations were performed for trajectories 1 through 3 (A--C) and 470,000 iterations for trajectory 4 (D). The initial values for parameters are marked by red circles. }}
\end{figure*}

Our next goal is choosing $M$ parameter sets, $\{\vec{\theta}_j\}_{m=1}^{M}$, in such a way that they are independent of each other. These parameter sets cannot be randomly chosen from the above sampling trajectories because the different samples may be highly correlated. To estimate decorrelation time of parameter values so as to obtain independent samples, we analyzed the autocorrelation function of parameter values between the 100,000th and 200,000th iteration steps (Figure~S2).
For most parameters, the autocorrelation function quickly decays to zero --- e.g., $K_{M,ON}$, $k_{cat,ON}$, and $k_{TAI}$.
On the other hand, for some parameters such as $K_{M,OFF}$ and $k_{cat,OFF}$, the parameter values did not converge and slowly drifted over time, possibly because they do not impact the cost function in a significant way.
For these two parameters, which were poorly constrained in the histogram analysis, the autocorrelation function did not completely decay even after 50,000 steps.  Nevertheless, the signs of autocorrelation function at time lags beyond 30,000 steps were roughly random --- it could be positive or negative depending on the sampling ranges. Thus, we used 30,000 iterations as the decorrelation time.

\begin{figure*}[h!!]
\centerline{\epsfig{figure=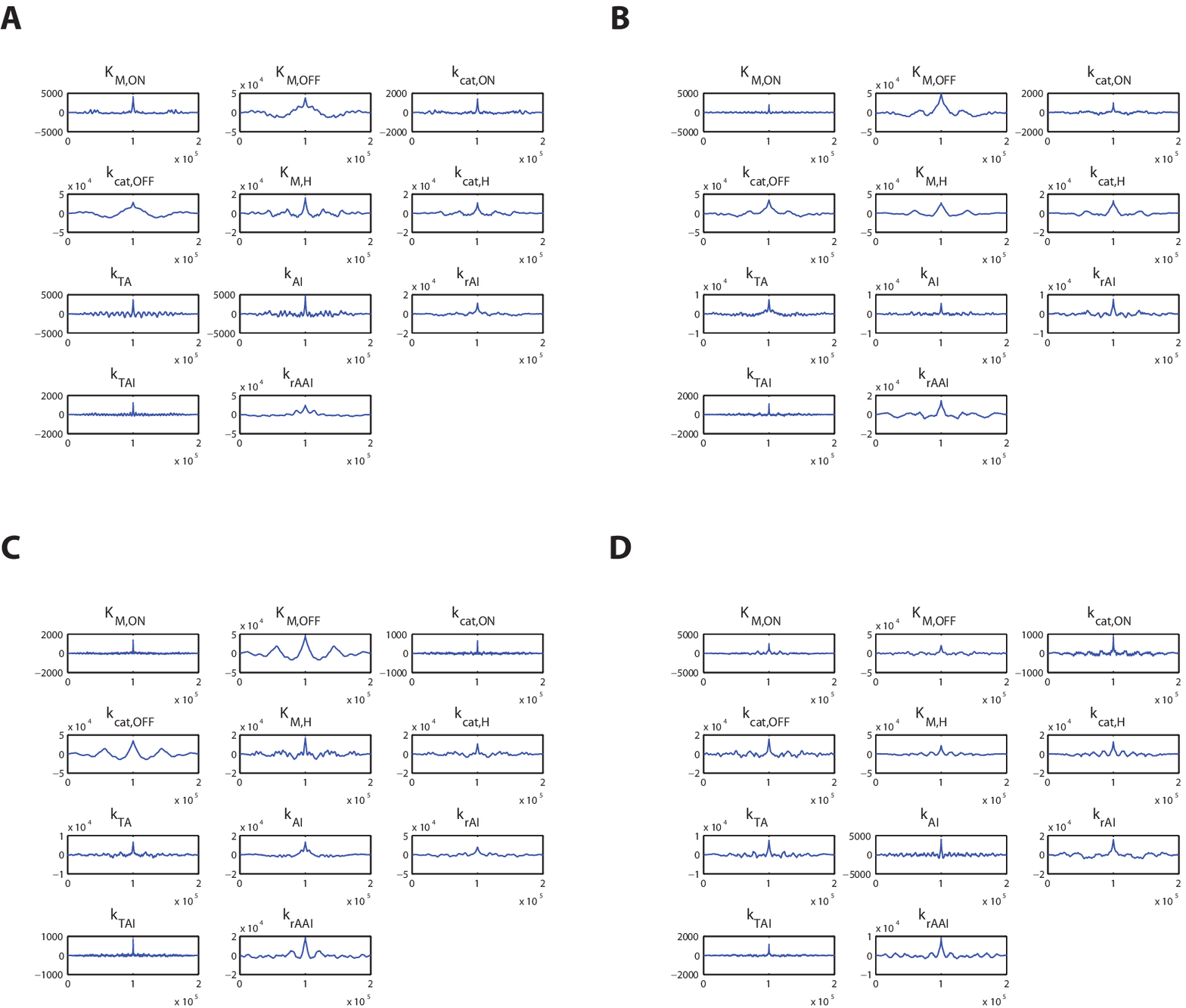,width=.95\textwidth}}
\caption{\textsf{Autocorrelation functions for parameter-sampling trajectories. ({\textbf{A--D}}) The log-scale parameter values ($\vec{\theta}$) between the 100,000th and 200,000th iteration steps for the sampling trajectories are plotted as autocorrelation functions. Autocorrelation functions are for the sampling trajectory 1 (A), 2 (B), 3 (C), and 4 (D). Decorrelation time for each parameter can be estimated by measuring lag times beyond which the autocorrelation function is close to zero.}}
\end{figure*}

For sampling trajectories 1, 2, and 3 (with 620,000 iterations each), we discarded the first 20\% of iterations and selected one parameter set per 30,000 iterations, resulting in 17 parameter sets per trajectory. This collection of parameter sets --- 51 in total ($M$ = 51) --- was used for simulation and compared with the experimental results in Figure~6. The line plots in Figure~6 show the central 95\% of predictions --- the two parameter sets with lowest rA output predictions and the two parameter sets with highest rA output predictions were omitted --- i.e., the four parameter sets with largest errors.  For the final 51 parameter sets, the sum squared errors ranged from 2.03 to 2.43 $\mu$M$^2$ for the whole training set ($N$ = 50 measurements).  One can construct an empirical covariance matrix ${\boldsymbol\theta}$ from the ensemble of parameters $\{\vec{\theta}_j\}_{m=1}^{M}$,
\[ {\boldsymbol\theta} = \langle (\vec{\theta}- \langle \vec{\theta} \rangle ) (\vec{\theta}- \langle \vec{\theta} \rangle )^T \rangle ,
\]
where the angle brackets denote ensemble average.  An eigenvalue decomposition of this matrix (called principal component analysis (PCA)) can then be inverted and information about soft and stiff modes are obtained analogous to that using the Hessian because PCA Hessian $P$ = ${\boldsymbol\theta}^{-1}$. Mode spectrum and eigenvector projections are shown with eigenvalue-eigenvector correspondence indicated by the numbers 1--11 (Figure~S3).

\begin{figure*}[h!!]
\centerline{\epsfig{figure=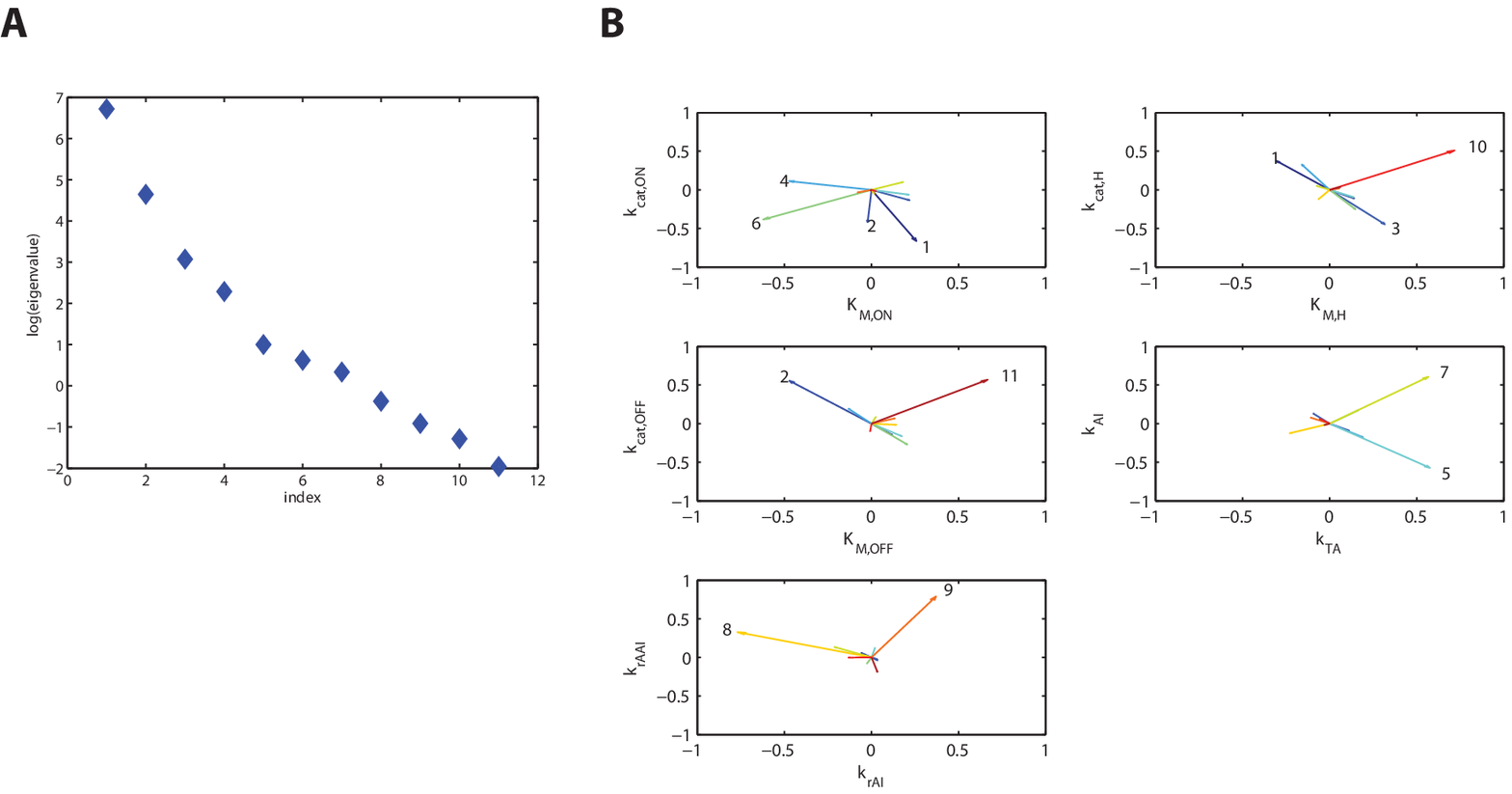,width=.95\textwidth}}
\caption{\textsf{Mode spectrum and eigenvector projections from PCA of inverse of covariance matrix ${\boldsymbol\theta}$. (\textbf{A}) Eigenvalues plotted in log scale. The first few eigenvalues are much larger than the rest --- i.e., the few combinations of parameters associated with large eigenvalues can be used to explain and fit most of the variation in the experimental and simulation data.
(\textbf{B}) Eigenvector projections. Eigenvector-eigenvalue correspondence is indicated by the numbers 1--11: small numbers correspond to large eigenvalues (stiff modes) and large numbers correspond to small eigenvalues (soft modes).}}
\end{figure*}

These results show that the eigenvalue spacing is almost uniform in logspace; there is no clear cutoff between ``stiff'' and ``soft''  eigenvalues. However, the wide spacing between eigenvalues indicate that the stiffest few parameter combinations can be used to explain and fit most of the variations in the experimental data.  These ``sloppy model'' features have been observed in many other high-dimensional multiparameter nonlinear models~\cite{Brown2003}. These characteristics are typically encountered in models for biochemical regulatory networks in biological organisms~\cite{gutenkunst2007universally}.

Let us proceed to interpret stiff modes found in the eigenvector projections. The variance in the stiffest mode (mode 1) was largely explained by contributions from $K_{M,ON}$ and $k_{cat,ON}$ as well as $K_{M,H}$ and $k_{cat,H}$. The angle on $K_{M,ON}$-$k_{cat,ON}$ plane indicates that these two values must change in opposite directions, which would change the ON-state switch transcription rate significantly.  At the same time, the angle on $K_{M,H}$-$k_{cat,H}$ plane indicates that these two values also must change in opposite directions, which in this case would change the degradation rate significantly.  Together, the signs for these four parameters in eigenvector indicates that the stiffest mode is captured by an increased ON-state transcription rate with a decreased degradation rate and vice versa --- i.e., ``production/degradation balance''.
The second stiff mode (mode 2) was dominated by contributions from $K_{M,OFF}$ and $k_{cat,OFF}$, indicating that the changes in opposite directions for these two parameters would result in significant changes in the OFF-state switch transcription rate --- i.e., ``leakiness''.
The inhibition rate $k_{TAI}$ contributed significantly in modes 3 and 4. However, since the parameters that change in the same or the opposite directions in relation to $k_{TAI}$ were not clearly identified, the eigenvalue projections involving $k_{TAI}$ are not shown in this figure.

Now we examine some of the soft modes revealed in the eigenvector projections.  The variance in the softest mode (mode 11) is mostly captured by parameters $K_{M,OFF}$ and $k_{cat,OFF}$.  The angle on $K_{M,OFF}$-$k_{cat,OFF}$ plane indicates that these two values must change together to preserve the OFF-state transcription rate, $k_{cat,OFF}$/$K_{M,OFF}$.  Similarly, the next soft mode (mode 10) is captured by parameters $K_{M,H}$ and $k_{cat,H}$, preserving the degradation rate, $k_{cat,H}$/$K_{M,H}$. The projections onto $k_{TA}$-$k_{AI}$ plane revealed that the softer mode (mode 7) corresponds to concerted changes in the activation rate ($k_{TA}$) and the annihilation rate ($k_{AI}$), which would change the time-scale of switching but preserve the average switch states. The $k_{rAI}$-$k_{rAAI}$ plane analogously associated the softer mode (mode 9) with the concerted changes of these two parameters.

\end{document}